\documentclass[twocolumn,times]{aastex631}

\usepackage{comment}

\shorttitle{CO(2--1)/CO(1--0) of GMCs in NGC~1300}
\shortauthors{Maeda et al.}
\graphicspath{{./}{figures/}}
\begin{document}

\title{CO(2--1)/CO(1--0) line ratio on $\sim$100 parsec scale in the nearby barred galaxy NGC~1300}

\author[0000-0002-8868-1255]{Fumiya Maeda}
\affiliation{Institute of Astronomy, Graduate School of Science, The University of Tokyo, 2-21-1 Osawa, Mitaka, Tokyo 181-0015, Japan}
\affiliation{Department of Astronomy, Kyoto University Sakyo-ku, Kyoto 606-8502, Japan}

\author[0000-0002-1639-1515]{Fumi Egusa}
\affiliation{Institute of Astronomy, Graduate School of Science, The University of Tokyo, 2-21-1 Osawa, Mitaka, Tokyo 181-0015, Japan}

%\collaboration{6}{(AAS Journals Data Editors)}

\author[0000-0003-3844-1517]{Kouji Ohta}
\affiliation{Department of Astronomy, Kyoto University Sakyo-ku, Kyoto 606-8502, Japan}

\author[0000-0002-2107-1460]{Yusuke Fujimoto}
\affiliation{Earth and Planets Laboratory, Carnegie Institution for Science, 5241 Broad Branch Road, NW, Washington, DC 20015, USA}

\author{Asao Habe}
\affiliation{Graduate School of Science, Hokkaido University, Kita 10 Nishi 8, Kita-ku, Sapporo, Hokkaido 060-0810, Japan}

\author[0000-0003-3983-5438]{Yoshihisa Asada}
\affiliation{Department of Astronomy, Kyoto University Sakyo-ku, Kyoto 606-8502, Japan}



\begin{abstract}

CO(2--1) emission is often used as a tracer of the giant molecular clouds (GMCs) as an alternative to CO(1--0) emission in recent years. Therefore, understanding the environmental dependence of the line ratio of  CO(2--1)/CO(1--0), $R_{21}$, on GMC scale is important to accurately estimate the mass of the GMCs. We thus measured the $R_{21}$ in the strongly barred galaxy NGC~1300, where star formation activity strongly depends on galactic structure, on $\sim 100$ pc scale. CO images were obtained from ALMA and Nobeyama 45-m telescope. The resultant typical $R_{21}$ in NGC~1300 is $0.57 \pm 0.06$. We find environmental variations in $R_{21}$; it is the highest in the bar-end region ($0.72 \pm 0.08$), followed by arm ($0.60 \pm 0.07$) and bar regions ($0.50 \pm 0.06$). GMCs with H$\alpha$ emission show a systematically higher ratio ($0.67 \pm 0.07$) than those without H$\alpha$ ($0.47 \pm 0.05$). In the bar region, where massive star formation is suppressed, H$\alpha$ emission is not associated with most GMCs, resulting in the lowest $R_{21}$. These results raise a possibility that properties of GMCs derived from CO(2--1) observations with the assumption of a constant $R_{21}$ are different from those derived from CO(1--0) observations. Furthermore, we find the $R_{21}$ measured on kpc scale tends to be lower than that of the GMCs probably due to the presence of an extended diffuse molecular gas in NGC~1300. 

\end{abstract}



\section{Introduction} \label{sec:intro}

Giant molecular clouds (GMCs) are crucial components for understanding galactic-scale star formation because they are the nurseries for stars.
The most abundant molecule in the GMCs is molecular hydrogen (H$_2$). However, as is well known, H$_2$ in the GMCs is hardly observable in emission. 
This is because H$_2$ has no permanent dipole moment and no corresponding dipolar rotational transitions. Furthermore, the lowest quadrupole rotational transition and vibrational transition of  H$_2$ are rarely observed due to the high excitation energy.
Unlike H$_2$, carbon monoxide (CO), which is the second most abundant molecule in the GMCs, has a permanent dipole moment. Since the moment of inertia of CO is much larger than that of H$_2$, 
CO has low excitation energy of the lowest rotational transition $J = $ 1--0 of $E/k \sim 5.5$~K, which is below typical GMC temperature \citep[$\sim 10-40$~K; e.g.,][]{scoville_h2_1987}.  Furthermore, critical density for CO(1--0) is reduced to be $\sim 200-300~\rm cm^{-3}$ by photon trapping due to its high optical depth, which is roughly comparable to the average density within the GMCs \citep{solomon_mass_1987}. Therefore, CO is easily excited to $J = 1$, and CO(1--0) is the most useful tracer of the bulk distribution of H$_2$ in the GMCs.

The properties of the GMCs have been investigated by CO(1--0) observations. In the Milky Way, molecular gas mass, size, and velocity dispersion of the GMCs are found to be typically 
$\sim 10^{4-6}~M_\odot$, $\sim 20 - 100$ pc, and $2 - 10~\rm km~s^{-1},$ respectively \citep[e.g.,][]{solomon_mass_1987,heyer_re-examining_2009,miville-deschenes_physical_2017}.
It has been reported that the properties of GMCs in the local group galaxies and nearby dwarf galaxies are 
similar to
those of the Milky Way; e.g.,
M31 \citep{rosolowsky_giant_2007},
M33 \citep{rosolowsky_giant_2003},
IC10 \citep{ohta_aperture_1992,leroy_molecular_2006}.
Further, extragalactic GMCs are detected on $\sim 50$ pc scale by CO(1--0) observations with Plateau de Bure Interferometer and Atacama Large Millimeter/submillimeter Array (ALMA), and environmental dependence of those properties are investigated; e.g., M51 \citep{colombo_pdbi_2014}, M83 \citep{hirota_alma_2018}, NGC~1300 \citep{maeda_properties_2020}.

To detect the GMCs, the higher excitation transition of CO(2--1) emission is sometimes used to achieve a high spatial resolution as an alternative to CO(1--0) \citep[e.g.,][]{gratier_giant_2012,utomo_giant_2015,wu_submillimeter_2017}. Furthermore, in recent years, CO(2--1) is often used in ALMA observations because CO(2--1) observations require much less time than CO(1--0)  to achieve the same mass sensitivity at the ALMA site. Using CO(2--1) data of a large sample of nearby galaxies \citep[e.g., PHANGS-ALMA;][]{leroy_phangs-alma_2021}, statistical studies of the properties of the GMCs have been conducted \citep[e.g.,][]{sun_cloud-scale_2018,rosolowsky_giant_2021}.
In such studies, molecular gas mass is often derived from CO(2--1) luminosity by assuming a constant CO(2--1)/CO(1--0) ratio (hereafter, $R_{21}$). The brightness temperature ratio of $R_{21} = 0.65 - 0.8$ is usually assumed based on  statistical studies of $R_{21}$ in nearby galaxies 
\citep[e.g.,][]{leroy_heracles_2009,leroy_molecular_2013,sandstrom_co--h2_2013,saintonge_xcold_2017,den_brok_new_2021}.

However, there are two problems with the assumption of constant $R_{21}$. The first is that $R_{21}$ can vary depending on environments in galaxies. 
The upper-level energy temperature 
and critical density of CO(2--1) emission are $E/k \sim 16.5 ~\rm K$ and $\sim 10^{3-4}~\rm cm^{-3}$, respectively, which are (slightly) higher than those of typical values of the GMCs. Therefore, $R_{21}$ can be influenced by the temperature and density of H$_2$ gas.
In the Milkly Way, the $R_{21}$ in the central region of $\sim 0.96$ is higher than the typical value in the Galactic disk of $0.6-0.7$, which is interpreted as higher temperature and density of H$_2$ gas in the central region \citep[e.g.,][]{sawada_tokyo-onsala-eso-calan_2001}. Such tendency for $R_{21}$ to be higher in central regions than in disks was also observed in nearby galaxies \citep[e.g.,][]{braine_co1-0_1992,leroy_heracles_2009,leroy_molecular_2013,yajima_co_2021}. The $R_{21}$ also changes inside the disk. In the Milky Way, $R_{21}$ varies from $\sim 0.75$ at 4 kpc to $\sim 0.6$ at 8 kpc in Galactocentric distance \citep{sakamoto_out--plane_1995,sakamoto_out--plane_1997}.
Furthermore, $R_{21}$ in inter-arm regions and bar regions, where star formation activity is low, tends to be lower than that in spiral arms and bar-end regions  \citep[e.g.,][]{koda_physical_2012,koda_systematic_2020,muraoka_co_2016,maeda_large_2020}.  According to the non-LTE analysis, gas in the interarm region should be colder and less dense than in the spiral arms \citep{koda_physical_2012}.

The second is that the assumption is based on the CO observations on kpc scale, not on GMC scale (i.e., $\leq 100$~pc). Molecular gas traced by CO(1--0) would consist of the GMCs and the extended diffuse molecular gas \citep[e.g.,][]{snow_diffuse_2006, liszt_imaging_2012, Shetty_2014}. It was reported that more than 50~\% of total molecular gas is such a diffuse gas which is distributed on scales larger than sub-kpc \citep[e.g.,][]{pety_plateau_2013,caldu-primo_spatially_2015,maeda_large_2020,Patra_2021}. Although the contribution of the diffuse gas is small in GMC observations, this component substantially contributes to the total CO flux in observations on kpc scale and can lower the $R_{21}$ due to its low density.
Therefore,  the $R_{21}$ can depend on the scale (beam size), and thus measurements of $R_{21}$ on GMC scale are indispensable. In the Milky Way, $R_{21}$ measurements on GMC scale or a scale that can resolve GMCs have been made \citep[e.g.,][]{sakamoto_large_1994,sakamoto_out--plane_1997,oka_co_1996,seta_enhanced_1998,nishimura_revealing_2015}. However, there are only a few examples of $R_{21}$ measurements on GMC scale in nearby galaxies; e.g., LMC \citep{sorai_co_2001}, M33 \citep{druard_iram_2014}, and NGC~628 \citep{herrera_headlight_2020}.

\begin{figure}[t!]
\begin{center}
\includegraphics[width=\hsize]{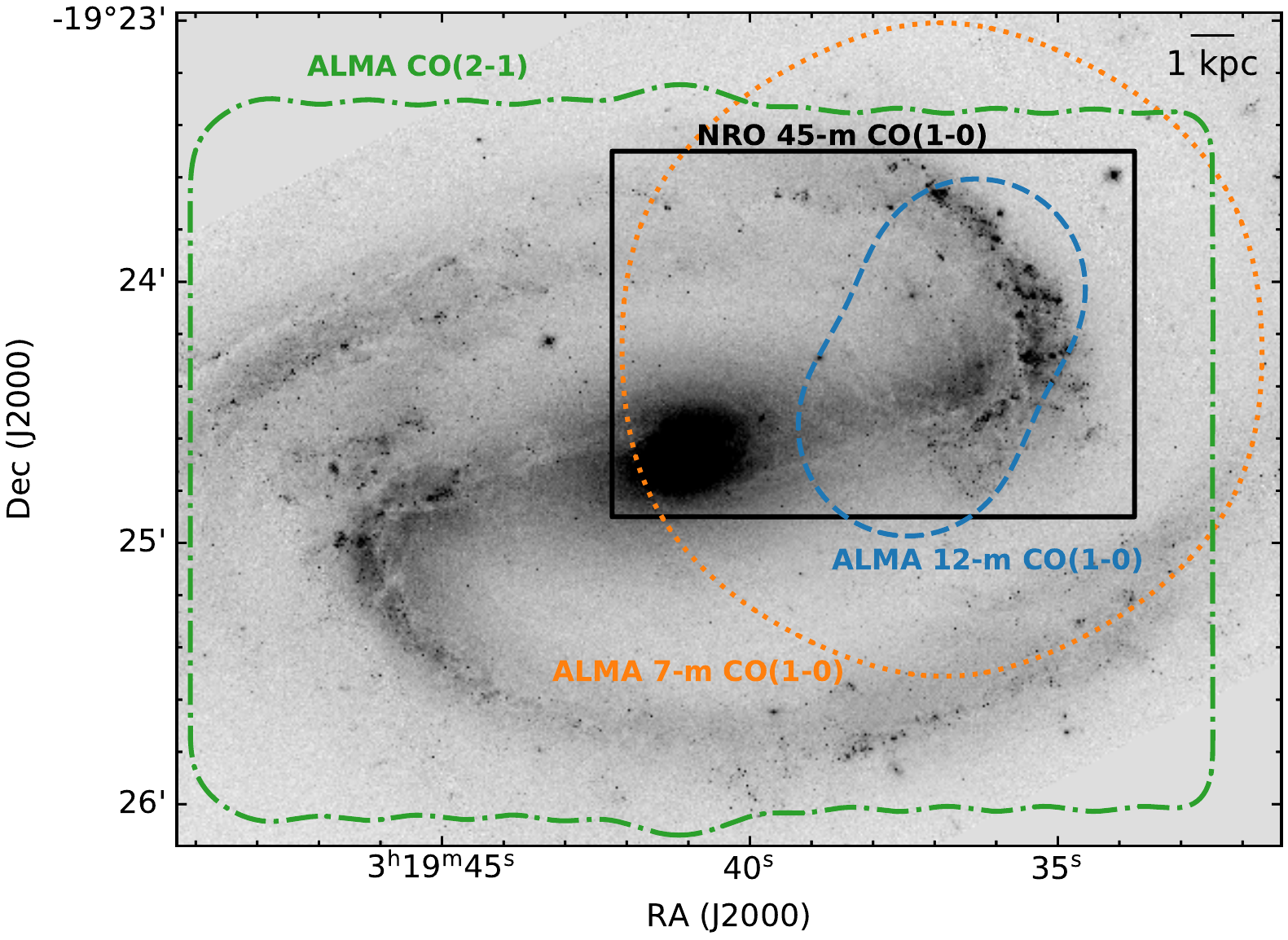}
\caption{$V$-band image of NGC~1300 taken with F555W filter on Advanced Camera for Surveys (ACS) of the Hubble Space Telescope (HST). We obtained this image from the Hubble Legacy Archive (HLA; \url{https://hla.stsci.edu/}). The FoVs of CO(1--0) observations with ALMA 12-m array, 7-m array, and NRO 45-m telescope are represented as blue dashed line, orange dotted line, and black solid line, respectively. The green dash-dotted represents the FoV of CO(2--1) observations with ALMA.
}
\label{fig:FoV}
\end{center}
\end{figure}

For these reasons, it is urgent to investigate the environmental dependence of $R_{21}$ on GMC scale.
In this paper, we present $R_{21}$ in the nearby strongly barred galaxy NGC~1300 (Figure~\ref{fig:FoV}) on $\sim 100$ pc scale using ALMA and the 45-m single-dish telescope of Nobeyama Radio Observatory (NRO\footnote{The Nobeyama Radio Observatory (NRO) is a branch of the National Astronomical Observatory of Japan, National Institutes of Natural Sciences.}).
This is the first study of $R_{21}$ on $\sim100$~pc scale in a nearby barred galaxy.
In NGC~1300, unlike the arm regions, star formation activity in the bar region is suppressed despite the presence of GMCs \citep{maeda_properties_2020}. Since such  significant differences in star formation activity are seen, NGC~1300 is thought to be an ideal laboratory to unveil the impact of the galactic environment on the $R_{21}$. The $R_{21}$ in NGC~1300 was already measured at 1.67 kpc resolution by \citet{maeda_large_2020} and a  systematic variation was found; in the bar-end region, the $R_{21}$ is the highest ($\sim 0.7$), followed by the arm region ($\sim 0.4 - 0.6$) and the bar region ($\sim0.2 - 0.4$). 
Our main goals are to (1) measure the $R_{21}$ on $\sim 100$ pc scale, and (2) investigate its environmental dependence, (3) compare with the previous lower resolution results, and (4) investigate the influence of the assumption of constant $R_{21}$ on the study of the GMC properties.

There are two main methods to investigate the properties of the molecular gas in the high-resolution CO image. One is 
the cloud identification approach, which was commonly used in previous GMC studies \citep[e.g.,][]{williams_determining_1994, rosolowsky_biasfree_2006}. Although this method is useful for identifying isolated structures, clumps finding algorithm requires a number of tuning parameters and assumptions, which are often not physically motivated. The other is the pixel-by-pixel approach, which has been often used in recent years \citep[e.g.,][]{2012ApJ...759L..26S,leroy_portrait_2016,egusa_molecular_2018,sun_cloud-scale_2018}. This approach is nonparametric and can treat all detected emissions. However, this approach does not provide information on the spatial extent of the molecular gas structure. In this study, we measure the $R_{21}$ on $\sim100$~pc scale based on both methods to minimize systematics.

This paper is structured as follows: In Section~2, we describe our CO(1--0) observations with ALMA and NRO 45-m. We also summarize archival CO(2--1) and H$\alpha$ data in this section. Then, Section 3 presents the method and result of the $R_{21}$ measurements based on the pixel-by-pixel method. Section~4 is the same as  Section~3, but based on GMC identification. Discussions and summary are given in Sections~5 and 6, respectively.
We adopt the distance of NGC~1300 to be 20.7~Mpc, calculated from the systemic velocity with corrections for the Virgo cluster, the Great Attractor, and the Shapley concentration of $1511~\rm km~s^{-1}$ \citep{mould_thehubble_2000} and the Hubble constant of $73~\rm km~s^{-1}~Mpc^{-1}$.

\section{Observations and Data Reduction}
\subsection{CO(1--0)}

We made a $^{12}$CO(1--0) cube combining the data obtained with the 12-m array of ALMA, 7-m array of the Morita Atacama Compact Array (ACA) of ALMA, and the NRO 45-m. The ALMA 12-m array observations were carried out as an ALMA Cycle 5 program (ID = 2017.1.00248.S, PI = F.Maeda) on 2017 December 30 and 2018 January 6, 7, 13, and 16 with a total on-source time of $\sim 5$ hrs. Since the details of the observations are described by \citet{maeda_properties_2020}, a brief summary is given here. Two pointings were set in order to cover the western bar, arm, and bar-end regions; the field of view (FoV) within the primary beam correction factor smaller than 2.0 is shown as a blue dashed line in Figure~\ref{fig:FoV}. About 44 antennas were used with C43-5 configuration in which the projected baseline length ranged from 15.1~m to 2.5~km, corresponding to a maximum recoverable scale (MRS) of $\sim$21.4 arcsec at 115 GHz. Bandpass and phase were calibrated with J0423–0120 and J0340–2119, respectively. 
The Precipitable Water Vapor (PWV) was typically $1-5$ mm during the observations.

The ACA 7-m array observations were performed in ALMA Cycle 7 (program ID = 2019.2.00139.S, PI = F.Maeda) as a follow-up observation program to obtain short-spacing data for the region observed by the 12-m array. To cover the FoV of the 12-m array observations, seven pointings were set. The FoV within the primary beam correction factor smaller than 2.0 is shown as a orange dotted line in Figure~\ref{fig:FoV}.  The observations were conducted in ten execution blocks, divided into 7 periods (2020 January 14, 15, 16, 25, March 3, 6, 8, and 14), resulting in a total on-source time of about 8 hours. Most of these observations were performed with 11 antennas. The projected baseline length ranged from 8.9~m to 28.2~m, corresponding to a MRS of $\sim$57.6 arcsec at 115 GHz. We used the Band 3 receiver with the central frequency of 114.669 GHz, channel width of 244.1 kHz ($\sim 0.64~\rm km~s^{-1}$), and bandwidth of 500.0 MHz ($\sim 1300~\rm km~s^{-1}$). This set-up was almost the same as the 12-m array observations. Bandpass was calibrated with J0334-4008, J0423-0120, J0519-4546, and J0522-3627. Phase were calibrated with J0348-1610. The PWV was typically $5-8$ mm during the observations.

To include zero-spacing data, we used the data obtained with the NRO 45-m single dish telescope. Since the details of the observations are described by \citet{maeda_large_2020}, a brief summary is given here. The observations were carried out on 2019 February 16, 17, 18, and 20. The observed area is shown as a black rectangle in Figure~\ref{fig:FoV}. We performed the on-the-fly mapping using two sets of scans along the right ascension and declination of the rectangle for a total on--source time of $\sim15$ hrs.  We used the multi-beam receiver, FOur-beam REceiver System (FOREST) with the central frequency of 114.699 GHz, channel width of 488.28 kHz ($\sim 1.3~\rm km~s^{-1}$), and bandwidth of 1~GHz ($\sim 2600~\rm km~s^{-1}$) at 115~GHz.
The effective angular resolution was $\sim 16.7$ arcsec. The line intensity was calibrated by the chopper wheel method. The pointing accuracy was kept within 3 arcsec by observing an SiO maser source. 

Calibrations of raw visibility data of the 12-m and 7-m array were done using the Common Astronomy Software Applications (\textsc{CASA}) ver. 5.1.1. and 5.6.1., respectively, and the observatory-provided calibration scripts. Imaging was made with \textsc{casa} ver. 5.7.2.
First, we concatenated the two calibrated-visibility data sets using the \textsc{casa} task \verb|concat|. Then, we reconstructed the image using the \verb|multiscale | CLEAN algorithm with Briggs weighting with robust of 0.5. We generated a CO(1--0) data cube limiting the velocity range between $1100 - 2100~\rm km~s^{-1}$. We chose a channel width and pixel size of $5.0~{\rm km~s^{-1}}$ and 0.12 arcsec, respectively, which are the same as those by \citet{maeda_properties_2020}. The resulting cube was imaged in the FoV of the 12-m array observations.

The NRO 45-m observation data were analysed using the same method as \citet{maeda_large_2020}, but we smoothed the spectrum by binning to $5.0~\rm km~s^{-1}$ here not to $20.0~\rm km~s^{-1}$ as in the previous paper.
We combined the interferometric and single-dish data with the \textsc{casa} task \verb|feather|. The resultant angular resolution is 0.44 arcsec $\times$ 0.30 arcsec, corresponding to 44 pc $\times$ 30 pc, with a postion angle of $-80.2^\circ$. The resultant rms noise of the combined image without primary beam correction is $0.54~\rm mJy~beam^{-1}$, corresponding to 380~mK. We confirmed that the combined image with spatially smoothing to the beamsize of the NRO 45-m reproduces the total flux measured with the NRO 45-m data alone.

\begin{figure*}[htbp]
\begin{center}
\includegraphics[width=\hsize]{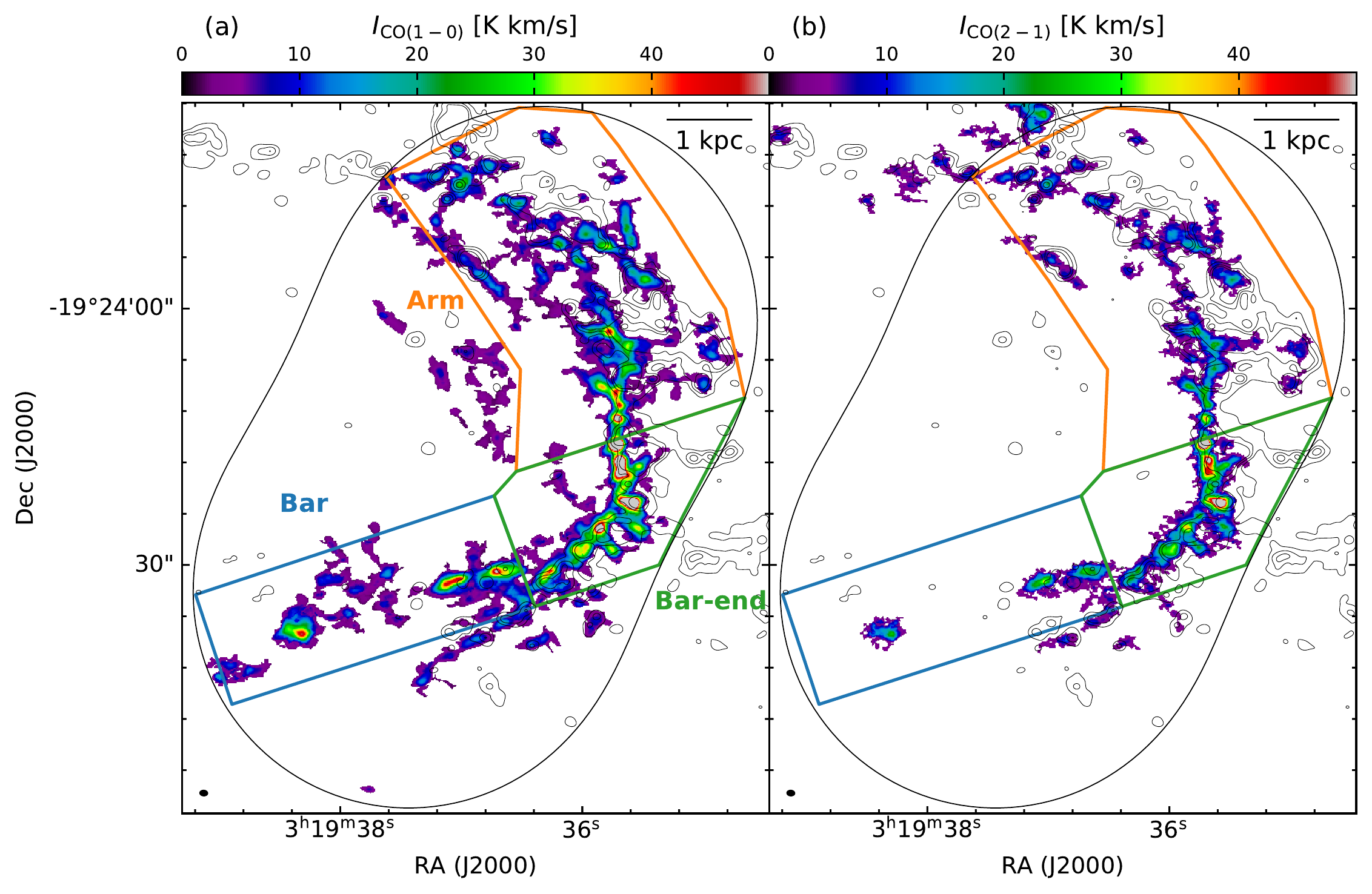}
\caption{(a) Velocity-integrated intensity map of CO(1--0) in NGC~1300. 
A large oblong line represents the FoV of the CO(1--0) observation with ALMA 12-m array.
The color solid lines indicate the environmental mask defined by \citet{maeda_properties_2020}.
{\it Bar}, {\it arm}, and {\it bar-end} are indicated with blue, orange, and green lines, respectively. 
The black filled circle at the lower left corner represents the beam size of $1.04^{\prime\prime} \times 0.79^{\prime\prime}$. 
Black contours show H$\alpha$ emission with 3$\sigma$, 10$\sigma$, and 20$\sigma$, respectively. The CO missions are identified by CPROPS with the parameters of THRESHOLD = 4.0 and EDGE = 1.5 (see Section~\ref{sec: GMC identification}).(b) Same as panel (a), but for CO(2--1). }
\label{fig:0th moment maps}
\end{center}
\end{figure*}

\subsection{Archival data}
\subsubsection{CO(2--1)}
We made a $^{12}$CO(2--1) cube using the archival data which was observed with 12-m array and ACA (7-m + Total Power) under project 2018.1.01651.S (PI = A. Leroy) and 2015.1.00925.S (PI = B. Guillermo), respectively. These observations were part of PHANGS-ALMA project \citep{leroy_phangs-alma_2021}. This project mapped
the whole disk of 90 nearby massive star-forming galaxies in CO(2--1) at an angular resolution of about 1 arcsec. The FoV within the primary beam correction factor smaller than 2.0 is shown as a green dash-dotted line in Figure~\ref{fig:FoV}. The band and channel widths were set to be about 1.0 GHz ($\sim 1300~\rm km~s^{-1}$) and 244.141 kHz ($\sim 0.32~\rm km~s^{-1}$) for each observation. For the 12-m and 7-m array observations, we calibrated raw visibility data using \textsc{casa} and the observatory-provided calibration scripts. After concatenating the two calibrated-visibility data sets, we reconstructed the image using the \verb|multiscale| CLEAN algorithm with Briggs weighting with robust of 0.5. The velocity range, channel width, and pixel size were set to be the same as the CO(1--0) cube. Then, using the \textsc{casa} task \verb|feather|, the interferometric cube
was combined with the cube obtained with the Total Power.
The resultant angular resolution is $1.04^{\prime\prime} \times 0.79^{\prime\prime}$, corresponding to 104~pc $\times$ 79~pc, with a postion angle of 86.4$^\circ$. The rms noise of the data cube is $4.02~\rm mJy~beam^{-1}$, corresponding to 114~mK.

\subsubsection{H$\alpha$}

The $R_{21}$ can be affected by star formation activity.
In the observations on kpc scale resolutions, the $R_{21}$ tends to be correlated with star formation rate (SFR) \citep[e.g.,][]{koda_physical_2012, yajima_co_2021}. 
The high $R_{21}$ in high SFR region is interpreted that OB stars are heating molecular gas (i.e., stellar feedback), or/and star formation are occurring as a result of dense molecular gas. To investigate the relation between star formation and $R_{21}$ at GMC scale, we use H$\alpha$ emission, which is one of the tracers of massive star formation (i.e., H\textsc{ii} region). We use a continuum-subtracted H$\alpha$ image of NGC~1300 obtained by \citet{maeda_large_2020}. They made the image using archival images taken with a broad-band F555W filter and a narrow-band H$\alpha$ filter, F658N, on the ACS of the {\it HST}. By comparing the two images in the region where H$\alpha$ emission is not seen, they determined the underlying stellar continuum in F658N image (see Section 2.3.2 in the paper for the detail).

\begin{figure*}[htbp]
\begin{center}
\includegraphics[width=\hsize]{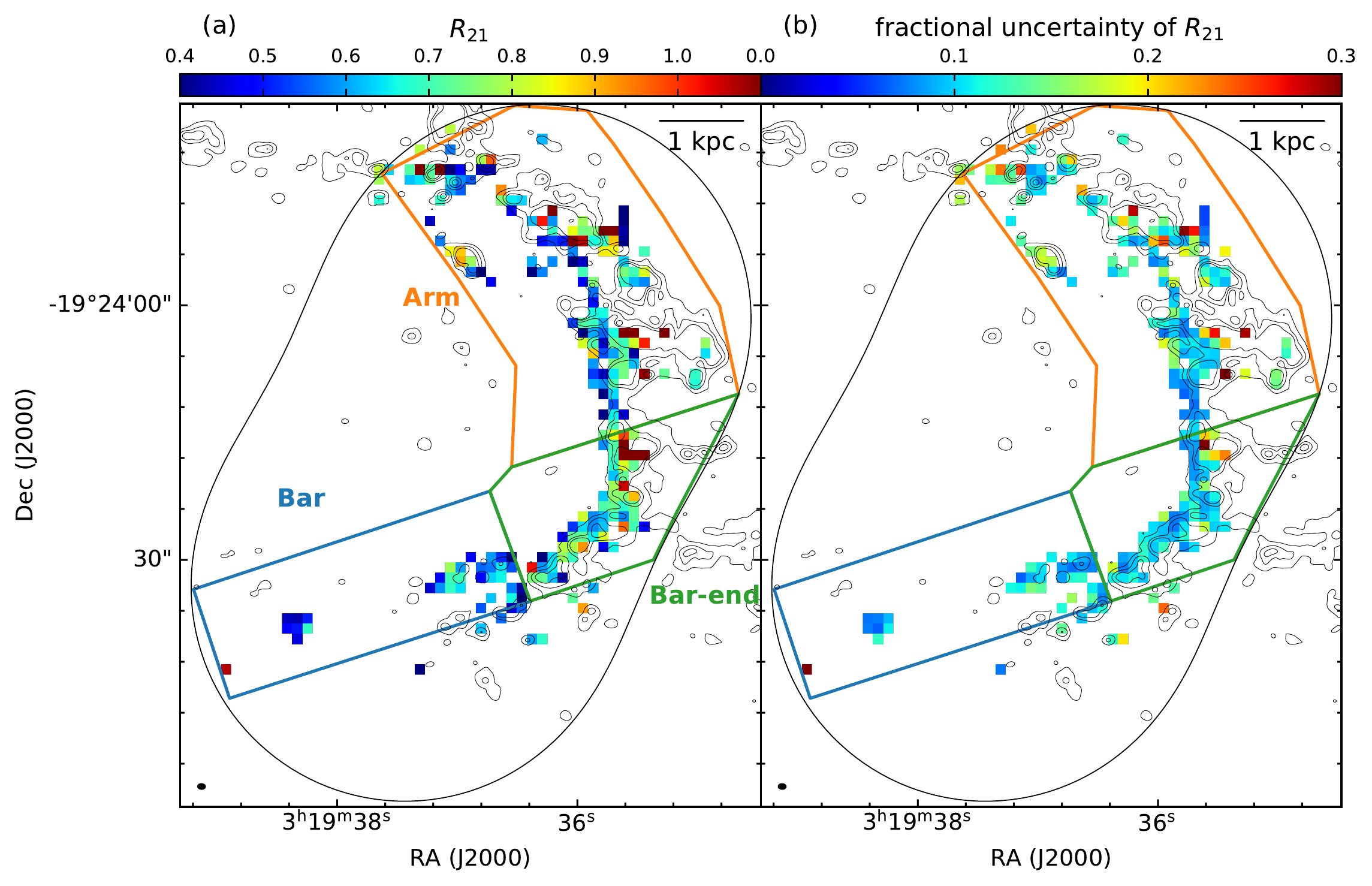}
\caption{Same as Figure~\ref{fig:0th moment maps}, but for $R_{21}$ (a) and its fractional uncertainty. Pixel size is 120~pc.}
\label{fig:R21 pix-by-pix map}
\end{center}
\end{figure*}

\subsection{Homogenization and comparison of CO and H$\alpha$ maps} \label{sec: homogenization}
To match the spatial resolutions, the CO(1--0) cube was convolved to the beam size of the CO(2--1) cube of $1.04^{\prime\prime} \times 0.79^{\prime\prime}$ by using the \textsc{casa} task imsmooth.
The rms noise of the convolved CO(1--0) image is $3.75 ~\rm mJy~beam^{-1}$, corresponding to 98~mK. This is comparable to that of the CO(2--1) cube. Similarly, the H$\alpha$ image was regrided to the pixel scale of the CO images and convolved to the beam size of the CO(2--1) cube.

Figures \ref{fig:0th moment maps}(a) and (b) show velocity-integrated intensities maps (i.e., moment-zero map) of CO(1--0) and CO(2--1), respectively.
We used CPROPS \citep{rosolowsky_biasfree_2006} to identify CO emission lines in the cubes, and pixels identified as so-called "island" are included to make these moment maps (see Section~\ref{sec: GMC identification} for more detail).
Black contours show H$\alpha$ emission with 3$\sigma$, 10$\sigma$, and 20$\sigma$, respectively.
As shown in this figure, CO(2--1) tends to be detected in H$\alpha$ bright regions, while CO(1--0) is detected in both H$\alpha$ dark and bright regions.
Note that CO emission is not detected downstream of the H$\alpha$ bright regions. We will discuss the evolution in physical conditions of the molecular gas in Section~\ref{sec: evolution in physical conditions of the molecular gas}.
In this paper, we use environmental masks defined by \citet{maeda_properties_2020}. In Figure~\ref{fig:0th moment maps}, the definitions of {\it bar}, {\it arm}, and {\it bar-end} regions are indicated with blue, red, and green polygons, respectively. Each region is determined by referring to the 0th moment map of CO(1--0) and the $V$-band image. The {\it bar} region covers the dark lane and associated spurs that are connected almost perpendicularly to the dark lane. The {\it bar-end} region covers the intersection region of the bar and arm.

\section{Pixel-by-pixel $R_{21}$ }
\subsection{Measurement} \label{sec: pix-by-pix measurement}

In this section, we present the line ratio $R_{21}$ based on a 
pixel-by-pixel analysis. 
As the spatial resolution of the data cube is $\sim 1^{\prime\prime}$ and the pixel size is $0.^{\prime\prime}12$, the data cube is oversampled in the spatial direction. Therefore, we rebinned the data cube to the pixel size of $1.^{\prime\prime}20$, corresponding to 120~pc. This scale is comparable to the radius of the identified GMC described in Section~\ref{sec: R21 of GMCs}. Using the rebinned CO(1--0) and CO(2--1) cubes, we measured $R_{21}$ in the spectrum of each pixel (line-of-sight) as follows; we first identified the consecutive channels in which signals are above $3\sigma_{\rm rms}$ within the velocity range of $1500 - 1750~\rm km~s^{-1}$, where significant CO emissions were detected in previous studies \citep[e.g.,][]{maeda_large_2020}. Here, the $\sigma_{\rm rms}$ is calculated in each line-of-sight by using line-free channels. Then, we expanded these channels to include all adjacent channels in which signals are above $1.5\sigma_{\rm rms}$. Note that the method of identifying emission lines here is independent of the one in Section~\ref{sec: homogenization}.
The velocity-integrated intensity of each pixel is defined as the sum of the identified channels. Finally, we measured $R_{21}$ for pixels in which both CO(1--0) and CO(2--1) emissions were detected. Here, we excluded one pixel in which the difference in the intensity-weighted mean velocity of both emissions was $\sim 35 ~\rm km~s^{-1}$ ($=$~7~channels) as a false detection. Note that the difference is less than $13~\rm km~s^{-1}$ ($<$~3~channels) for all other pixels. 

We estimated the observation error of the velocity-integrated intensity as $\sigma = \sqrt{N} \sigma_{\rm rms} \Delta V$, where $N$ is the number of channels used in the integration, and  $\Delta V$ is the channel width of $5.0~\rm km~s^{-1}$, respectively.
Additionally, we considered the absolute flux calibration accuracy of $\pm 5~\%$ and $\pm 10~\%$ of CO(1--0) in ALMA Band~3 and CO(2--1) in Band~6, respectively (ALMA Technical Handbook).

\begin{deluxetable}{cccc}
\tablecaption{$R_{21}$ in NGC~1300 by pix-by-pix method. \label{tab:R21 pix-by-pix}}
\tablewidth{0pt}
\tablehead{Region & w/  H$\alpha$ & w/o  H$\alpha$ & both}
\decimalcolnumbers
\startdata
Whole   & $0.69^{+0.09}_{-0.09}$ & $0.59^{+0.11}_{-0.11}$ & $0.65^{+0.10}_{-0.09}$ \\
Bar     & $0.52^{+0.13}_{-0.04}$ & $0.57^{+0.07}_{-0.09}$ & $0.57^{+0.08}_{-0.09}$ \\
Arm     & $0.68^{+0.08}_{-0.10}$ & $0.58^{+0.12}_{-0.14}$ & $0.65^{+0.10}_{-0.10}$ \\
Bar-end & $0.73^{+0.08}_{-0.08}$ & $0.63^{+0.09}_{-0.12}$ & $0.72^{+0.07}_{-0.10}$ \\
\enddata
\tablecomments{$R_{21}$ is noted as $M^{D75}_{D25}$, where $M$, $D25$, and $D75$ are the median, the distance to the 25th 
percentile from the median, and the distance to the 75th percentile from the median of the number distribution, respectively.}
\end{deluxetable}

\begin{figure*}[htbp]
\begin{center}
\includegraphics[width=\hsize]{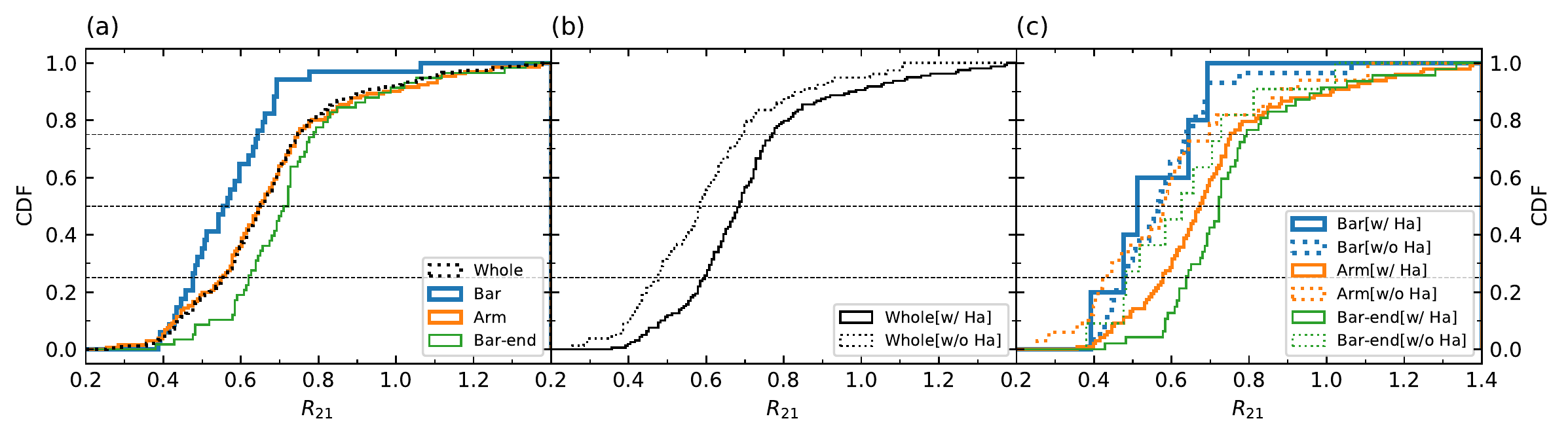}
\caption{$R_{21}$ measured by pixel-by-pixel method. (a) Normalized cumulative distribution function of the $R_{21}$ in the whole region and each environment. (b) Same as panel (a), but for pixels with H$\alpha$ and without H$\alpha$ in the whole region.
(c) Same as panel (b), but for each environment.}
\label{fig:R21 pix-by-pix hist}

\includegraphics[width=\hsize]{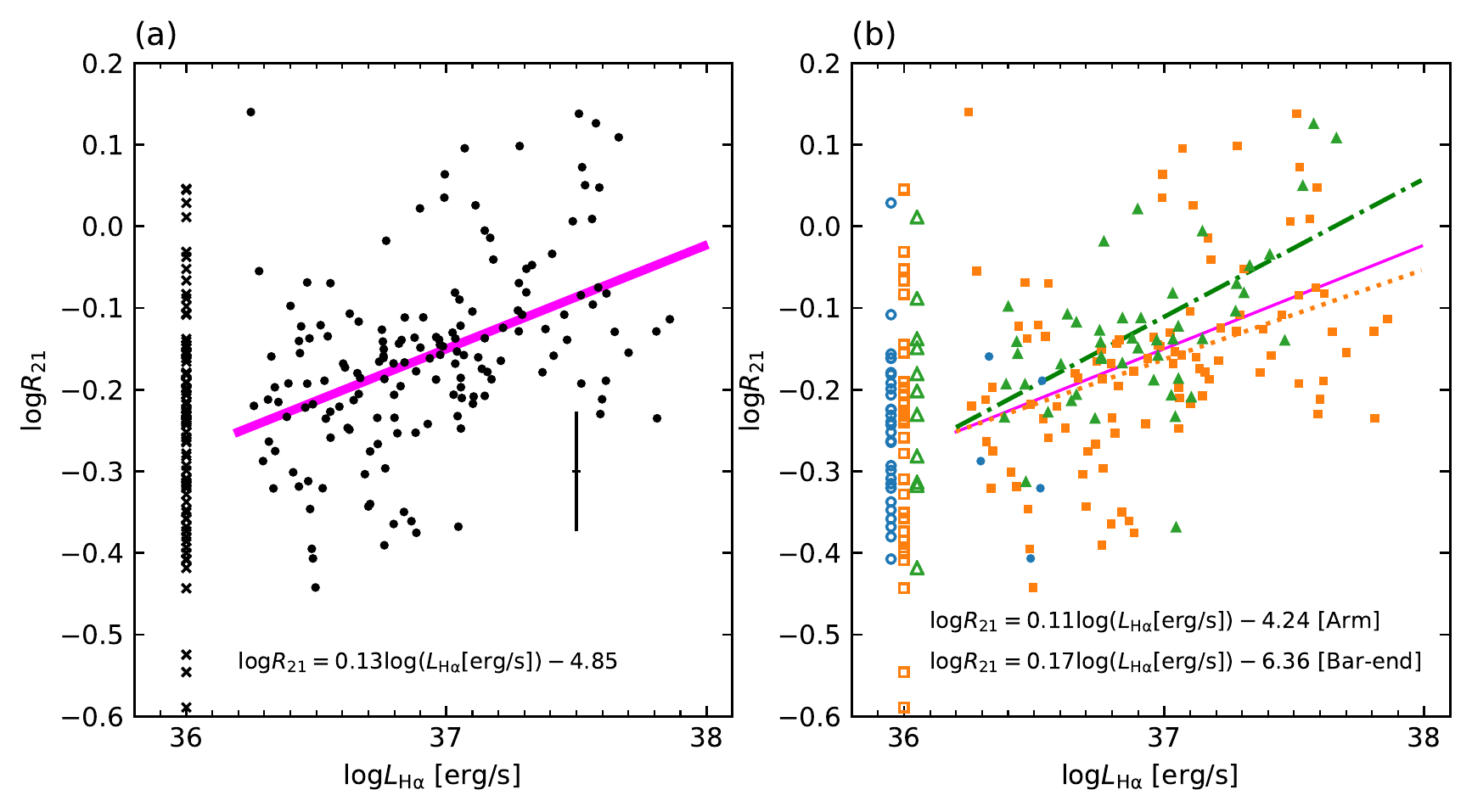}
\caption{(a) Relationship between $L_{\rm H\alpha}$ and $R_{21}$ in the whole region.  Black filled circles show the pixel with H$\alpha$. We identified pixels above 3$\sigma$ level ($1.75 \times 10^{36}~\rm erg~s^{-1}$) as significant H$\alpha$ emissions. The median error bar is indicated as a cross. Magenta solid line shows the best-fitting line and its slope is given in the bottom of the panel.
Black x marks show the pixel without H$\alpha$ and are plotted at $L_{\rm H\alpha} = 10^{36}~\rm erg~s^{-1}$ for convenience.
(b) Same as panel (a), but for the pixels in {\it bar} (blue circle), {\it arm} (orange square), and {\it bar-end} (green triangle). Orange dotted line and green dash-dotted line show the best-fitting lines in {\it arm} and {\it bar-end}, respectively.}
\label{fig:R21_vs_Ha}
\end{center}
\end{figure*}

\begin{figure}[htbp]
\begin{center}
\includegraphics[width=\hsize]{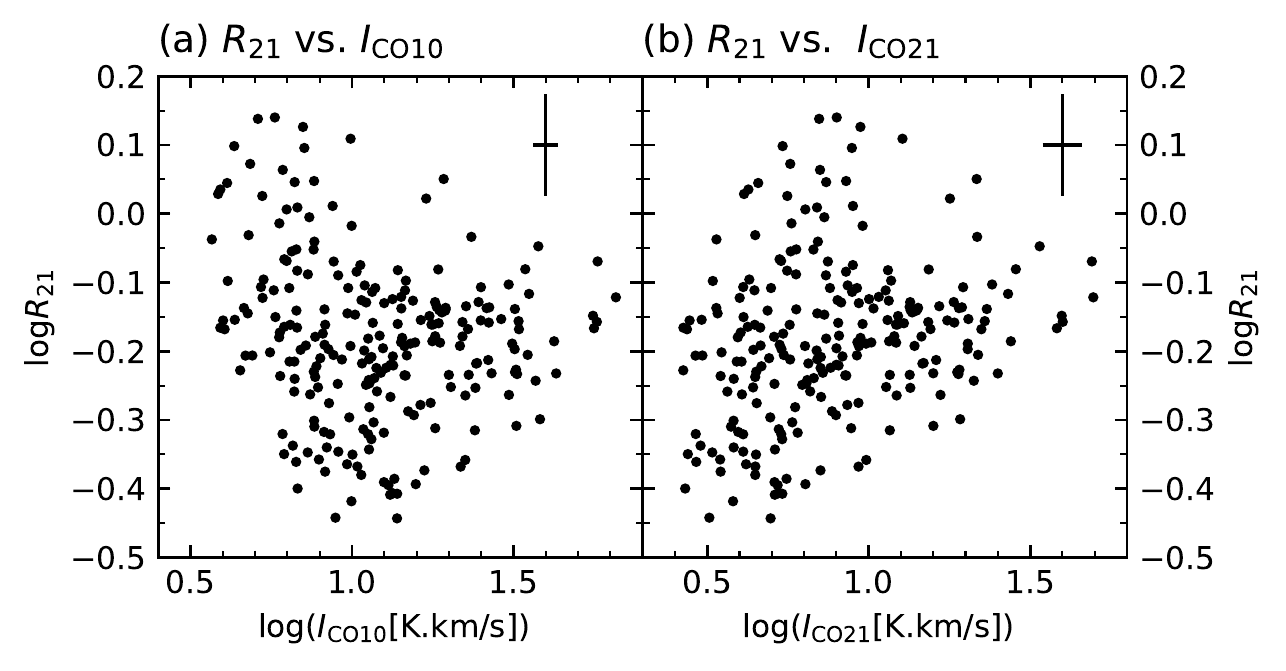}
\caption{(a) Relationship between CO(1--0) velocity-integrated intensity and $R_{21}$ of the pixels.
The median error bar is indicated as a black cross. 
(b) Same as panel (a), but for CO(2--1).}
\label{fig:R21_vs_Ico}
\end{center}
\end{figure}

\subsection{H$\alpha$ luminosity}

In this study, we investigate the relation between H$\alpha$ luminosity and $R_{21}$. The continuum-subtracted H$\alpha$ image is rebinned to the pixel size of $1.^{\prime\prime}20$.
As for the derivation of H$\alpha$ luminosity ($L_{\rm H_{\alpha}}$) from the H$\alpha$ flux density ($f_{\rm H_{\alpha}}$) of the continuum-subtracted H$\alpha$ image, we used the same calculation method as \citet{maeda_large_2020}. The $L_{\rm H_{\alpha}}$ is derived as 
\begin{equation}
    L_{\rm H_\alpha} = 4 \pi f_{\rm H_\alpha} W_{\rm eff} D^2 C_{\rm [N_{II}]} 10^{0.4 A_{\rm V}},
\end{equation}
where $W_{\rm eff}$ is the effective width of the F658N
filter of 74.75~\AA, $D$ is the distance to NGC 1300 of 20.7~Mpc, $C_{\rm [N_{II}]}$ is
the correction factor to remove the [NII]
emission, and $A_{\rm V}$ is a correction for the
dust extinction. 
\citet{maeda_large_2020} estimated the [NII] contamination of 21~\% to $f_{\rm H_\alpha}$, or $C_{\rm [N_{II}]} = 0.79$, considering the flux density ratio of the [NII] doublet to H$\alpha$ and the transmission corrections.
The dust extinction was assumed to be $A_V = 1.0$~mag, which is a typical value for integrated galaxy discs obtained by \citet{leroy_estimating_2012}.
\citet{maeda_large_2020} compared star formation rate (SFR) derived from $L_{\rm H_\alpha}$ assuming $A_V =1.0$~mag and hybrid tracers (FUV + IR) at kpc scale and showed the difference was within a factor of 2 in H$\alpha$ bright regions. Therefore, the assumption of $A_V = 1.0$~mag is generally reasonable. We identified pixels above 3$\sigma$ level as significant H$\alpha$ emissions. 
Here, the rms noise of the flux density in the 120 pc pixel is $7.66~\times 10^{-20}~\rm erg~cm^{-2}~s^{-1}~\AA^{-1}$, corresponding to $5.83 \times 10^{35}~\rm erg~s^{-1}$. Among the pixels where both CO(1--0) and CO(2--1) are detected, the luminosity of the significant H$\alpha$ emission in the 120 pc pixel is $\log(L_{\rm H\alpha}/[\rm erg~s^{-1}]) = 36.89^{+0.27}_{-0.27}$, $36.49^{+0.04}_{-0.16}$ , $36.97^{+0.30}_{-0.31}$, and $36.91^{+0.16}_{-0.24}$ in the whole region, {\it bar}, {\it arm}, and {\it bar-end}, respectively.

\subsection{Results}

Figure~\ref{fig:R21 pix-by-pix map}(a) and (b) show the spatial distribution of derived $R_{21}$ and its fractional uncertainty, respectively. The number of pixels with a size of 120 pc in which both CO(1--0) and CO(2--1) emissions were detected is 239 in the whole region (34 in {\it bar}, 131 in {\it arm}, and 59 in {\it bar-end}, respectively). The median uncertainty is 20~\%. Figure~\ref{fig:R21 pix-by-pix hist}(a) shows the normalized cumulative distribution function of the $R_{21}$ in each environment. Figure~\ref{fig:R21 pix-by-pix hist}(b) and (c) show the ratio of the pixel with and without H$\alpha$. Results are also summarized in Table~\ref{tab:R21 pix-by-pix}.

The median $R_{21}$ in the whole region is 0.65 with a scatter of 0.19. Here, the scatter is defined as the distance to the 75th percentile from the 25th percentile (so-called "interquartile range", IQR). As shown in Figure~\ref{fig:R21 pix-by-pix hist}(a), we find an environmental difference in the $R_{21}$. The $R_{21}$ in {\it bar-end} is the highest (median of 0.72), followed by {\it arm} (0.65) and {\it bar} (0.57). The scatter is $0.17 - 0.20$.
We statistically checked the environmental variation of the $R_{21}$ distribution using the two-sided Kolmogorov--Smirnov (K–S) test. We used stats.ks\_2samp function of PYTHON’s Scipy package. As a result, the $p$-values ($p_{\rm KS}$) are low, confirming that the $R_{21}$ is statistically different among environments: the $p_{\rm KS}$ is $4.0 \times 10^{-3}$, $1.5 \times 10^{-3}$, and $3.4\times 10^{-2}$ for {\it bar} versus {\it arm}, {\it bar} versus {\it bar-end}, and {\it arm} versus {\it bar-end}, respectively.

$R_{21}$ differs depending on the presence or absence of H$\alpha$. In the whole region, the median $R_{21}$ of the pixels with H$\alpha$ of 0.69 is 0.10 higher than those without H$\alpha$ of 0.59.  The difference is also confirmed by $p_{\rm KS} = 4.4 \times 10^{-5}$. 
The $R_{21}$ of the pixels with H$\alpha$ in {\it bar-end} is the highest (median of 0.73), followed by {\it arm} (0.68) and {\it bar} (0.52). As for the pixels without H$\alpha$, while $R_{21}$ in {\it arm} and {\it bar} are comparable ($\sim 0.57$), $R_{21}$ in {\it bar-end} ($\sim 0.63$) tends to be slightly higher than those in {\it arm} and {\it bar}. 

Figure~\ref{fig:R21_vs_Ha} shows the relationship between the H$\alpha$ luminosity and the $R_{21}$ on $\sim 100$ pc scale. In panel (a), we find a moderate correlation for the pixels with H$\alpha$; the Spearman's rank correlation coefficient ($\rho_s$) is 0.46. The median $R_{21}$ for the pixels with $L_{\rm H\alpha} \leq 10^{37}~\rm erg~s^{-1}$ and $> 10^{37}~\rm erg~s^{-1}$ are 0.70 and 0.81, respectively. A magenta solid line represents the best-fitting line determined by the ordinary least-squares method: $R_{21} \propto L_{\rm H \alpha}^{0.13 \pm 0.20}$. Such a positive correlation has been seen in the observations on kpc scale resolutions \citep[e.g.,][]{koda_physical_2012,yajima_co_2021,Leroy_R2132_2021}.
\citet{Leroy_R2132_2021} show the relationship of $R_{21} \propto \Sigma_{\rm SFR}^{0.129}$, which is comparable to our results. This correlation is still seen if we divide the pixels into {\it arm} and {\it bar-end} (Figure~\ref{fig:R21_vs_Ha}(b)). The $\rho_s$ is 0.44 and 0.50 in {\it arm} and {\it bar-end}, respectively.  The slope in {\it bar-end } ($R_{21} \propto L_{\rm H \alpha}^{0.17 \pm 0.44}$) seems to be stepper than that in {\it arm} ($R_{21} \propto L_{\rm H \alpha}^{0.11 \pm 0.24}$). However, the significance of the difference is unclear because of the large scatter. Although it is unclear if there is a correlation in {\it bar} because of the small number of the data points, the H$\alpha$ luminosity  in {\it bar} is low ($\sim (2-3) \times 10^{36}~\rm erg~s^{-1}$) compared to {\it arm} and {\it bar-end}. This is the reason why the $R_{21}$ of the pixels with H$\alpha$ in {\it bar} shows lower values (Table~\ref{tab:R21 pix-by-pix}).

We also investigated relationships between the $R_{21}$ and velocity-integrated intensity of CO(1--0) and CO(2--1) emissions ($I_{\rm CO(1-0)}$ and $I_{\rm CO(2-1)}$). $I_{\rm CO(1-0)}$ is proportional to the molecular gas surface density. On kpc scale, $R_{21}$ increases weakly with the gas surface density \citep[e.g.,][]{koda_physical_2012, yajima_co_2021}. However, as shown in Figure~\ref{fig:R21_vs_Ico}(a), such a correlation is not seen in NGC~1300 on $\sim100$ pc scale ($\rho_s = -0.17$). While $R_{21}$ appears to increase weakly with  $I_{\rm CO(2-1)}$ (Figure~\ref{fig:R21_vs_Ico}(b)), the $\rho_s$ is low (0.31). Furthermore, no clear correlation was found between $R_{21}$ and line width of the CO emissions. The $\rho_s$ are $-0.27$ and $0.27$ for the CO(1--0) and CO(2--1), respectively. We confirmed that no clear correlation between $R_{21}$ and velocity-integrated intensity (or line width) of the CO emissions appears if we divide the pixels into each environment.

\begin{deluxetable*}{cccccccc}
\tablecaption{$R_{21}$ in NGC~1300 by stacking analysis \label{tab:stacking R21 pix-by-pix}}
\tablewidth{0pt}
\tablehead{ & \multicolumn{2}{c}{w/  H$\alpha$} & \multicolumn{2}{c}{w/o  H$\alpha$} & \multicolumn{3}{c}{both} \\
\colhead{Region} & $f_{\rm w/ CO21}$  & \colhead{$R_{21}$} & $f_{\rm w/ CO21}$ & \colhead{$R_{21}$} &  $f_{\rm w/ H\alpha}$ & $f_{\rm w/ CO21}$ & \colhead{$R_{21}$} 
}
\decimalcolnumbers
\startdata
Whole     & 78.8\% (160/203) & $0.67\pm0.07$ & 34.1\% (79/232) & $0.47\pm0.05$ & 46.7\% (203/435) & 54.9~\% (239/435) & $0.57\pm0.06$ \\
Bar     & 62.5\% (5/8)     & $0.59\pm0.08$ & 34.5\% (29/84)  & $0.49\pm0.06$ & 8.7\% (8/92)     & 37.0~\% (34/92) & $0.50\pm0.06$ \\
Arm     & 80.3\% (98/122)  & $0.66\pm0.07$ & 41.8\% (33/79)  & $0.48\pm0.06$ & 60.7\% (122/201) & 64.5~\% (131/201) & $0.60\pm0.07$ \\
Bar-end & 92.3\% (48/52)   & $0.73\pm0.08$ & 52.4\% (11/21)  & $0.61\pm0.08$ & 71.2\% (52/73)  & 80.8~\%(59/73) & $0.72\pm0.08$ \\
\enddata
\tablecomments{(2) Percentage of pixels with CO(2--1) out of pixels where CO(1--0) and H$\alpha$ were detected.
(4) Same as (2), but out of the pixels where CO(1--0) is detected but H$\alpha$ was not.
(6) Percentage of pixels with H$\alpha$ out of all pixels where CO(1--0) is detected.
(7) Same as (2), but out of all pixels where CO(1--0) is detected.}
\end{deluxetable*}

\subsection{Stacking analysis} \label{sec: Stacking analysis}

Since we used the pixels with significant CO(2--1) emissions, the above $R_{21}$ would be biased toward the CO(2--1) bright regions. In fact, we used only about half (239 pixels, 54.9~\%) of the total 435 pixels in which CO(1--0) was detected. In particular, as can be seen from Figure~\ref{fig:0th moment maps}(b), CO(2--1) tends to be weak in the region in which significant H$\alpha$ emission is not seen. Therefore, there is a possibility that the pixel-by-pixel method overestimate the  median $R_{21}$. In this section,  to assess the bias, we stack the CO spectra using the all pixels in which CO(1--0) was detected. 

\begin{figure}[htbp]
\begin{center}
\includegraphics[width=\hsize]{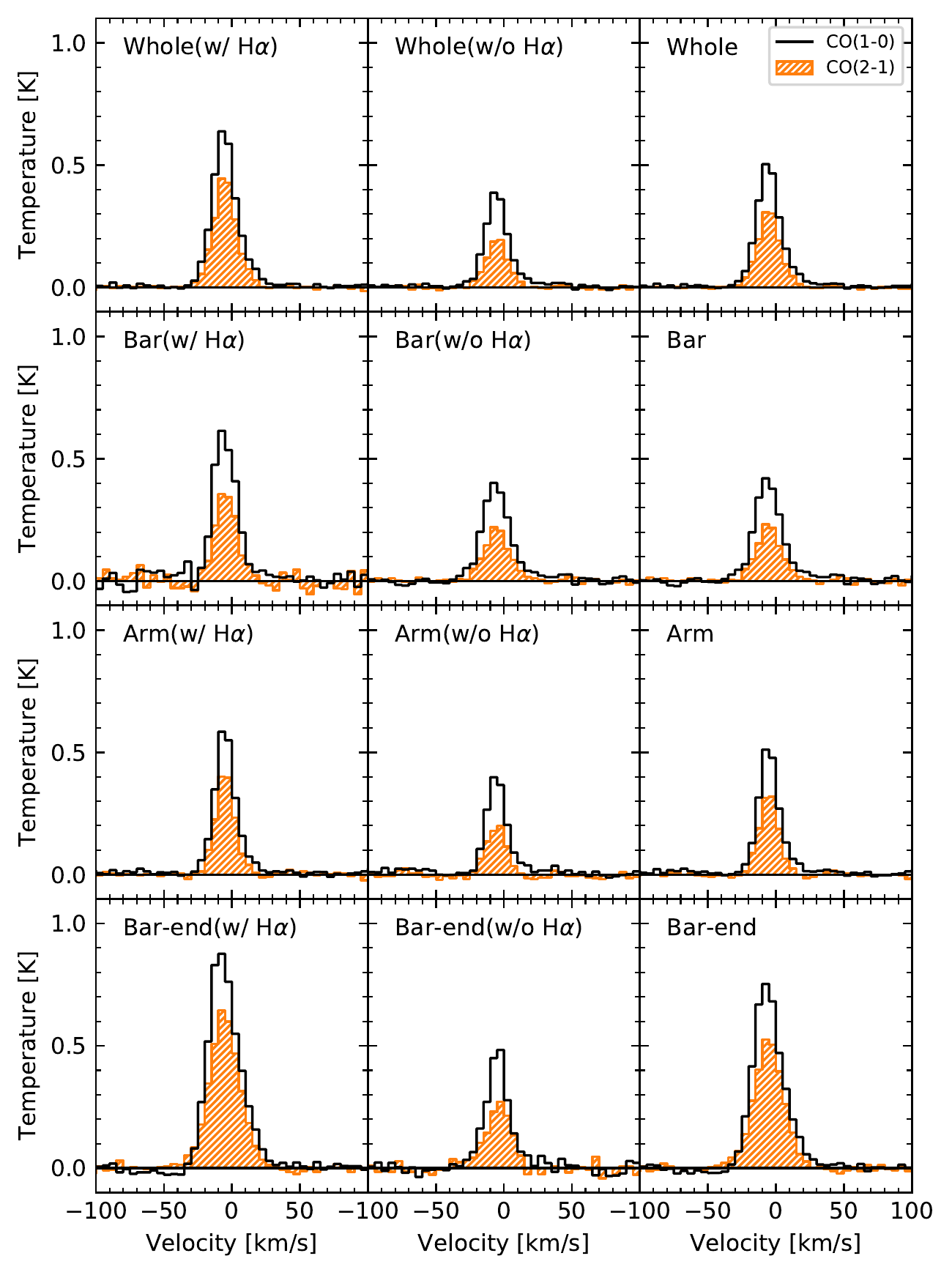}
\caption{Stacked CO spectra for each environment in NGC~1300. Black lines and orange hatched region indicate CO(1--0) and CO(2--1), respectively.
Left and middle columns show the results of stacking the pixels with and without H$\alpha$, respectively. Right column shows the results of stacking all pixels in which CO(1--0) was detected. 
}
\label{fig: stack profile}
\end{center}
\end{figure}

We perform a velocity-alignment stacking analysis originally proposed by \citet{schruba_molecular_2011,schruba_low_2012}. 
This method is commonly adopted for the CO spectra to measure the mean value in nearby galaxies \citep[e.g.,][]{morokuma-matsui_stacking_2015,muraoka_co_2016,yajima_co_2021}. For CO(1--0) emission, we stacked the spectra with velocity axis alignment based on the intensity-weighted mean velocity ($\bar{v}_{\rm CO(1-0)}$) calculated from each spectrum. For CO(2--1) emission, we stacked the spectra in the same method as CO(1--0).  For the spectra without CO(2--1) detection, we stacked the spectra based on the $\bar{v}_{\rm CO(1-0)}$. Finally, we calculated the $R_{21}$ using both stacked spectra in the same way as the pixel-by-pixel method (see Section~\ref{sec: pix-by-pix measurement}).
The absolute flux calibration accuracy mainly contributes to the uncertainty of the stacked  $R_{21}$. We also calculated the stacked $R_{21}$ by dividing the pixels into those with and without H$\alpha$. 
Fig.~\ref{fig: stack profile} shows the stacked CO spectra. 
Black lines and orange hatched region indicate CO(1--0) and CO(2--1), respectively. The resultant $R_{21}$ is listed in Table~\ref{tab:stacking R21 pix-by-pix}.

Here, we compare the $R_{21}$ derived by the pix-by-pix method with those by the stacking analysis (see also Figure~\ref{fig:R21 summary}). 
For the pixels with H$\alpha$, CO(2--1) was detected in most of the pixels: percentage of CO(2--1) detection ($f_{\rm w/ CO21}$) is high in all regions ($\sim 60-90~\%$). Thus, the stacked $R_{21}$ is comparable to or about 0.05 lower than the median of $R_{21}$ by the pixel-by-pixel method. On the other hand, for the pixels without H$\alpha$, $f_{\rm w/ CO21}$ is $\leq 50~\%$. The stacked $R_{21}$ is $0.08-0.12$ lower than the median of $R_{21}$ by the pixel-by-pixel method. That is, the pixel-by-pixel method is likely to overestimate the median $R_{21}$ of the pixels without H$\alpha$.

For all pixels in which CO(1--0) was detected, $f_{\rm w/ CO21}$ is 54.9, 37.1, 64.5, and 80.8~\% in the whole region, {\it bar}, {\it arm}, and {\it bar-end}, respectively. The pixel-by-pixel method is likely to overestimate the median $R_{21}$ in the whole region, {\it bar}, and {\it arm} where $f_{\rm w/ CO21}$ is not high; the stacked $R_{21}$ is $0.05-0.08$ lower than the median of $R_{21}$ by pixel-by-pixel method, while  both values are comparable in {\it bar-end}.
The cause for the large difference of 0.08 in the whole region is that the stacked $R_{21}$ includes not only the results in {\it bar} and {\it arm}, but also in the inter-arm region (i.e., outside the three environments) where $f_{\rm w/ CO21}$ is low ($21.7~\%$).

\begin{figure}[htbp]
\begin{center}
\includegraphics[width=\hsize]{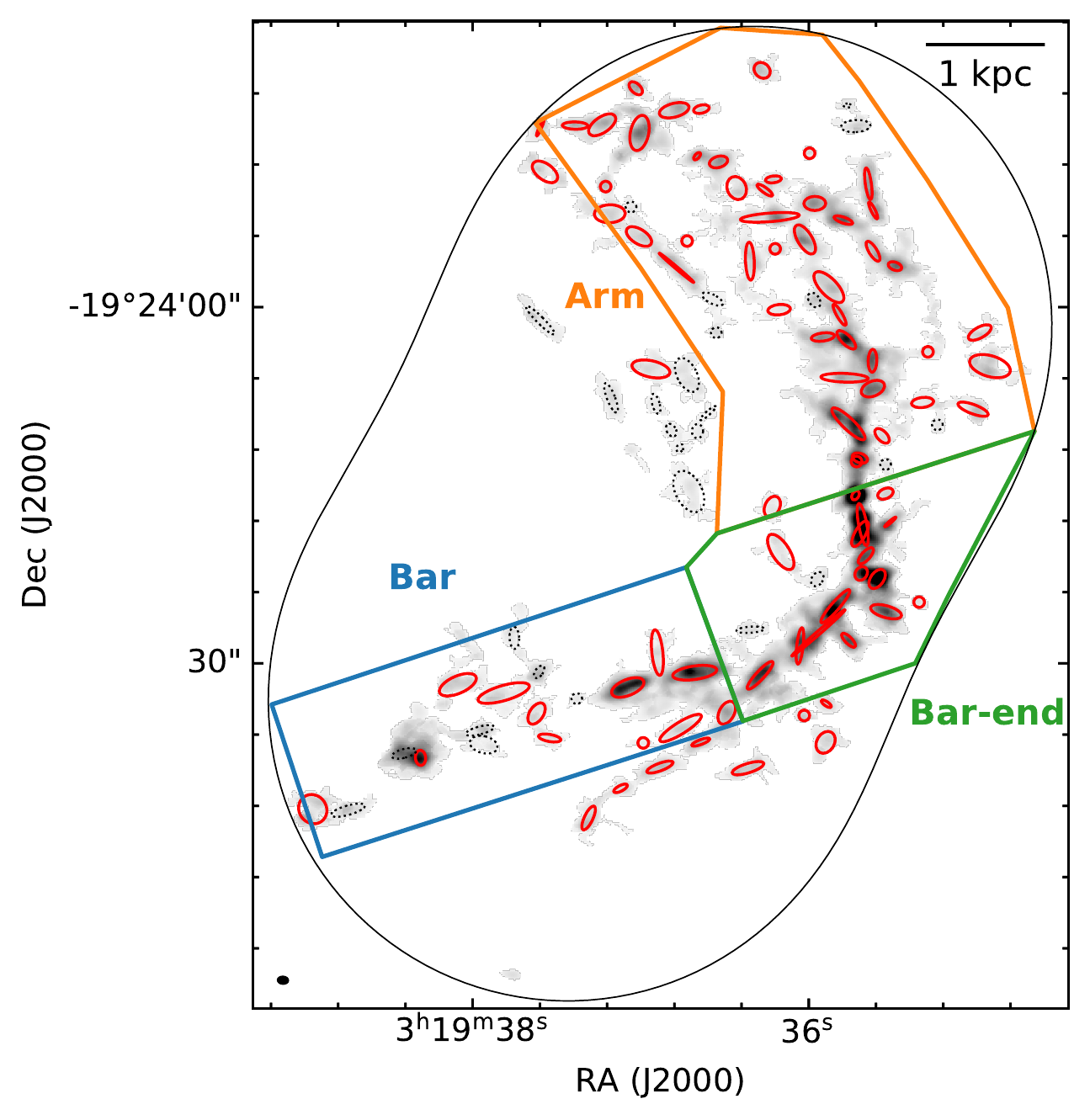}
\caption{
GMC distribution in NGC~1300 superimposed on the velocity-integrated CO(1--0) map (grey-scale). The GMCs are represented as ellipses. Red ellipses show the GMCs in which both CO(1--0) and CO(2--1) are detected. Black dotted ellipses show the GMCs in which only CO(1--0) is detected.
}
\label{fig:GMC_map}

\includegraphics[width=\hsize]{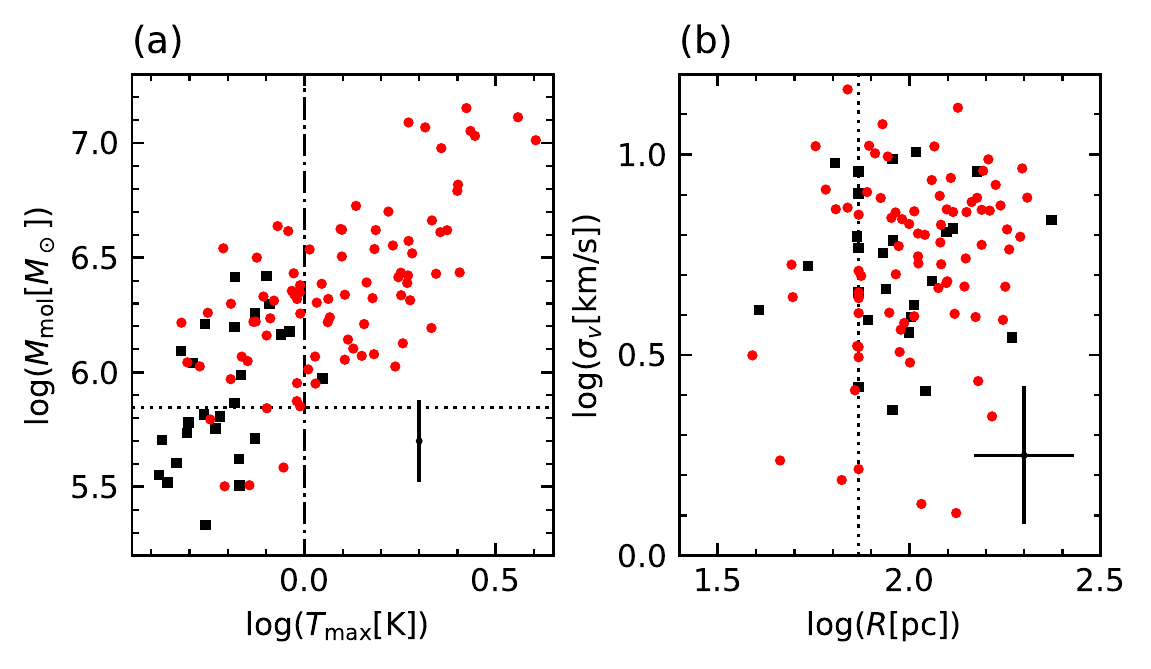}
\caption{(a) Relationship between the peak temperature and the molecular gas mass of the GMCs. Red filled circles show the GMCs in which $R_{21}$ is measured. Black filled squares show the GMCs in which  only CO(1--0) is detected.
Black dash-dotted line and dotted line represent  $T_{\rm max} = 1.0~\rm K$ and $M_{\rm mol} = 7.0 \times 10^5~M_\odot$, respectively, which show the CO(2--1) detectability limits (see text).
The median error of the molecular gas mass is indicated as a black bar.
(b) Relationship between the deconvolved radius and the velocity dispersion of the GMCs. Black dotted line represents $R = 73.7$~pc, corresponding to the beam size. For the unresolved GMCs, the radius is plotted in this line. The median error bar is indicated as a black cross.}
\label{fig:GMC_prop}
\end{center}
\end{figure}

\begin{deluxetable}{cccc}
\tablecaption{$R_{21}$ in NGC~1300 by GMC identification method. \label{tab:R21 GMC}}
\tablewidth{0pt}
\tablehead{Region & w/  H$\alpha$ & w/o  H$\alpha$ & both}
\decimalcolnumbers
\startdata
Whole     & $0.59^{+0.08}_{-0.05}$ & $0.46^{+0.12}_{-0.04}$& $0.58^{+0.08}_{-0.08}$\\
Bar     & $0.58^{+0.01}_{-0.07}$ & $0.43^{+0.10}_{-0.04}$ & $0.48^{+0.11}_{-0.09}$ \\
Arm     & $0.61^{+0.05}_{-0.07}$ & $0.48^{+0.12}_{-0.06}$ & $0.60^{+0.07}_{-0.07}$ \\
Bar-end & $0.68^{+0.04}_{-0.10}$ & -- & $0.68^{+0.04}_{-0.10}$ \\
\enddata
\tablecomments{Notation is the same as Table~\ref{tab:R21 pix-by-pix}}
\end{deluxetable}

\begin{figure*}[htbp]
\begin{center}
\includegraphics[width=\hsize]{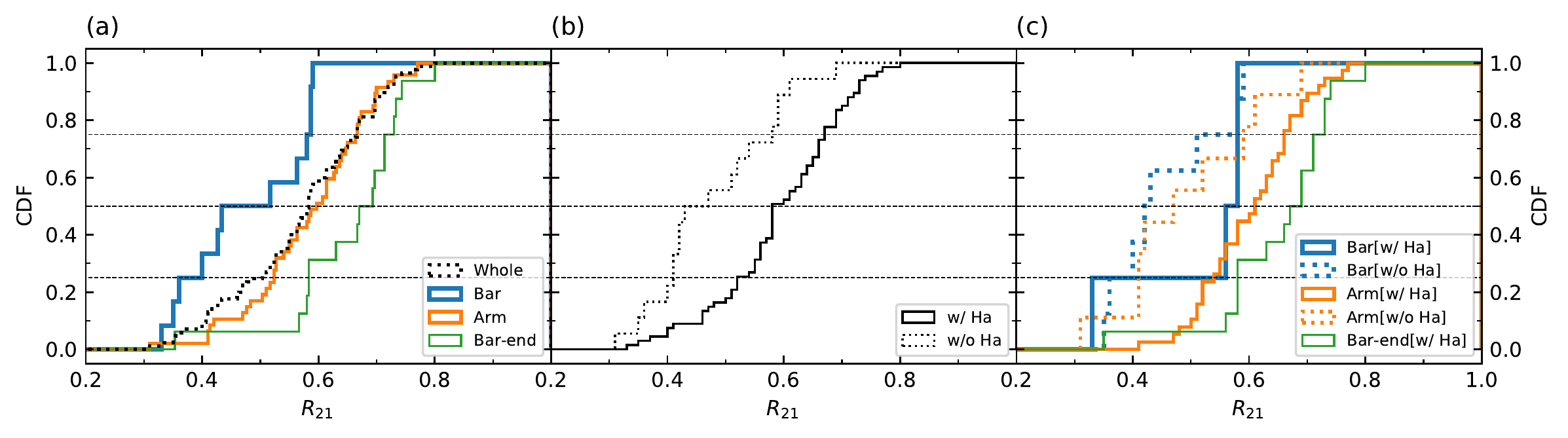}
\caption{$R_{21}$ measured by GMC identification method. (a) Normalized cumulative distribution function of the $R_{21}$ of the GMCs in the whole region and each environment. (b) Same as panel (a), but for GMCs with H$\alpha$ and without H$\alpha$ in the whole region.
(c) Same as panel (b), but for each environment.
In {\it bar-end}, all GMCs overlap with H$\alpha$ emission.}
\label{fig:R21 GMC hist}
\end{center}
\end{figure*}

\begin{figure}[htbp]
\begin{center}
\includegraphics[width=\hsize]{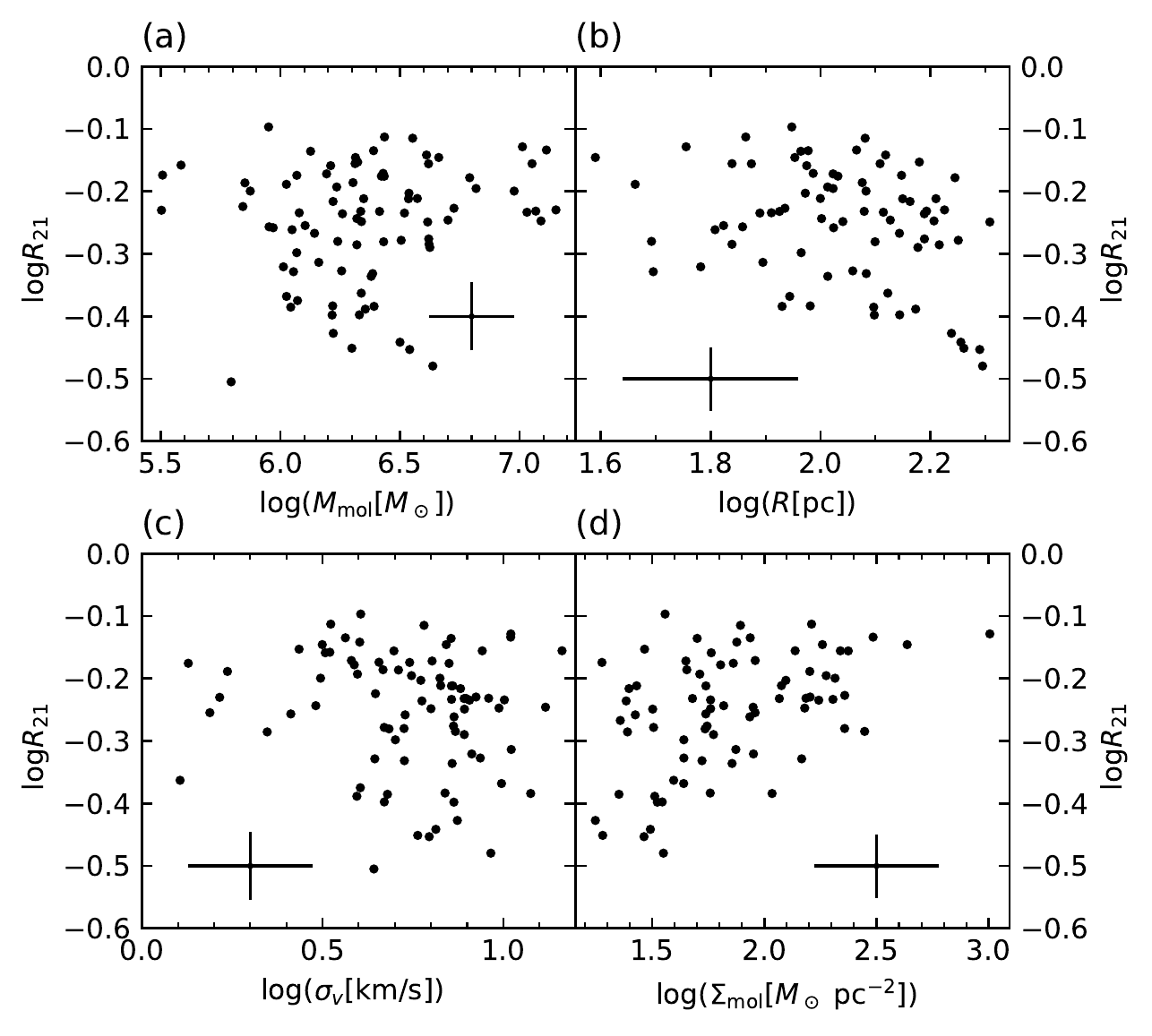}
\caption{Relationship between the $R_{21}$ and GMC properties: (a) molecular gas mass, (b) radius, (c) velocity deviation, and (d) molecular gas surface density. The median error bar is indicated as a black cross.}
\label{fig:R21 vs. GMC prop}
\end{center}
\end{figure}

\section{$R_{21}$ of GMCs} \label{sec: R21 of GMCs}
\subsection{Identification and Measurements} \label{sec: GMC identification}

In this section, we present the $R_{21}$ measured by identifying GMCs. We used the CO data cubes and H$\alpha$ image with the pixel size of $0.^{\prime\prime}12$. First, using 3D clumps finding algorithm \textsc{cprops} \citep{rosolowsky_biasfree_2006}, which is designed to identify GMCs well even at low sensitivities, we identified GMCs in the CO(1--0) cube. 

\textsc{cprops} begins with identification of regions with significant emissions within the data cube (so-called {\it islands}). \textsc{cprops} identifies pixels in which the signal is above $4.0 \sigma_{\rm rms}$ in at least two adjacent velocity channels, where $\sigma_{\rm rms}$ is the rms noise of the data cube. These {\it islands} are extended to include all adjacent pixels above $1.5 \sigma_{\rm rms}$. The parameters we used in \textsc{cprops} are \verb|THRESHOLD| $= 4.0$ and \verb|EDGE| $= 1.5$.  We called \verb|NONUNIFORM| flag because the noise in the CO(1--0) data cube is non-uniform due to the primary beam correction. Thus, the $\sigma_{\rm rms}$ is calculated for each line-of-sight. The velocity-integrated intensity map of the islands of the CO(1--0) emissions is shown in Figure~\ref{fig:0th moment maps}(a). Then the islands are divided into individual GMCs using a modified watershed algorithm. \textsc{cprops} searches for local maxima within a box of three times the beam and channel width. All local maxima are required to lie at least 2$\sigma_{\rm rms}$ above the merge level with another maximum. In this study, each local maximum is assigned to an individual independent cloud by setting \verb|SIGDISCONT| to be 0.0.  Note that these parameters in \textsc{cprops} are the same as those adopted in \citet{colombo_pdbi_2014} and \citet{maeda_properties_2020}.

A total of 111 GMCs were identified by \textsc{cprops}. Figure~\ref{fig:GMC_map} shows the spatial distribution of the GMCs. We identified 19, 55, and 18 GMCs in {\it bar}, {\it arm}, and {\it bar-end}, respectively. Based on \textsc{cprops} measurements, the radius ($R$),  velocity dispersion ($\sigma_v$), and molecular gas mass ($M_{\rm mol}$) of 111 GMCs range $105.4^{+33.9}_{-19.3}$~pc, $5.9^{+1.7}_{-1.8}~\rm km~s^{-1}$, and $1.8^{+1.1}_{-0.8} \times 10^6~M_\odot$, respectively. The $R$ and $\sigma_v$ are deconvolved by the beam and channel width, respectively. Although the CO-to-H$_2$ conversion factor, $\alpha_{\rm CO}$, may depend on environments as discussed Section \ref{sec: Impact of  using a constant R21}, we assumed the standard $\alpha_{\rm CO}$ of $4.4~M_\odot~\rm (K~km~s^{-1}~pc^2)^{-1}$ \citep{bolatto_co--h2_2013} here.

For each GMC, we measured the velocity-integrated CO(1--0) intensity as the sum of the intensities of the pixels identified as the GMC.  Note that \textsc{cprops} can correct for the sensitivity by extrapolating the flux to that we would expect to measure with perfect sensitivity (i.e., 0~K), but we did not make this correction here\footnote{The percentage of CO(1--0) luminosity contained in GMCs with respect to the total masked CO(1--0) luminosity is 41~\%.}. 
Then, in the CO(2--1) data cube, we extracted the pixels at the same position as the identified GMCs. We determined that CO(2--1) was detected in the GMC if there is at least one pixel with $S/N \geq 4$. CO(2--1) was detected in 85 of 111 GMCs shown as red ellipses in Figure~\ref{fig:GMC_map}
(12, 47, and 16 in {\it bar}, {\it arm}, and {\it bar-end}, respectively).
The sum of the CO(2--1) intensities in the pixels is defined as the velocity-integrated CO(2--1) intensity of the GMC.  Finally, we measured $R_{21}$ for the GMCs using both velocity-integrated intensities. Considering the rms noise and absolute flux calibration accuracy, the median uncertainty is 14~\%. Figure~\ref{fig:GMC_prop} shows the properties of the GMCs in which $R_{21}$ was measured. In this study, while CO(2--1) was detected in most of the GMCs with the peak temperature ($T_{\rm max}$) above 1.0~K, it was not detected in 43~\% of the GMCs with $T_{\rm max} \leq 1.0$~K. In particular, CO(2--1) was detected in only 28~\% GMCs with $M_{\rm mol} \leq 7.0 \times 10^5~M_\odot$. Conversely, $R_{21}$ was successfully determined for the most (86~\%) of the GMCs with $M_{\rm mol} > 7.0 \times 10^5~M_\odot$.

\subsection{Results}
Figure~\ref{fig:R21 GMC hist}(a) shows the normalized cumulative distribution function of the $R_{21}$ in each environment. Figure~\ref{fig:R21 GMC hist}(b) and (c) show the $R_{21}$ of the GMC with and without H$\alpha$. Results are also summarized in Table~\ref{tab:R21 GMC}. The median $R_{21}$ of the GMCs in the whole region is 0.58 with a scatter of 0.16. As with the results of the pixel-by-pixel method, environment differences in the $R_{21}$ are seen. The $R_{21}$ in {\it bar-end} is the highest (median of 0.68), followed by {\it arm} (0.60) and {\it bar} (0.48). The scatter is $0.14-0.20$. These differences are also confirmed by the K-S tests: the $p_{\rm KS}$ is $8.9 \times 10^{-3}$, $1.4 \times 10^{-3}$, and $3.5\times 10^{-2}$ for {\it bar} versus {\it arm}, {\it bar} versus {\it bar-end}, and {\it arm} versus {\it bar-end}, respectively. 

Significant H$\alpha$ emissions (i.e., above 3$\sigma$) overlap in 67 of the 85 GMCs. 
In {\it bar} and {\it arm},   4 (21~\%) and 38 (69~\%) GMCs  overlap with H$\alpha$ emission. In {\it bar-end}, all GMCs overlap with H$\alpha$ emission. As with the results of the pixel-by-pixel method, $R_{21}$ of the GMCs with H$\alpha$ tends to be higher than those without H$\alpha$. In the whole region, the median $R_{21}$ of the GMCs with H$\alpha$ of 0.59 is 0.13 higher than those without H$\alpha$ of 0.46 (Figure~\ref{fig:R21 GMC hist}(b); $\rho_{\rm KS} = 3.1 \times 10^{-3}$). 
The same trend can be seen in {\it arm} and {\it bar} (Figure~\ref{fig:R21 GMC hist}(c)).
As for the GMCs with H$\alpha$, the $R_{21}$ in {\it bar-end} is the highest ($0.58-0.72$), followed by {\it arm} ($0.54-0.66$) and {\it bar} ($0.51-0.59$). 
The above ratios are consistent with the $R_{21}$ by the stacking analysis (see also Figure~\ref{fig:R21 summary}). This would be because the pixels identified as GMCs in this method include those without significant CO(2--1) emission. 

As shown in Figure~\ref{fig:R21 vs. GMC prop}, we investigated the relationship between the $R_{21}$ and GMC properties: $M_{\rm mol}$, $R$, $\sigma_v$, and $\Sigma_{\rm mol}$. There is no clear correlation for $R_{21}$ vs. $M_{\rm mol}$ and $\sigma_v$: $\rho_s$ is $0.15$ and $-0.15$, respectively. The $R_{21}$ tends to slightly decrease with increasing $R$ ($\rho_s = -0.27$). As for $\Sigma_{\rm mol}$, we find a moderate correlation for $R_{21}$ vs. $\Sigma_{\rm mol}$: $\rho_s = 0.44$. The median $R_{21}$ for the GMCs with $\Sigma_{\rm mol} \leq 100~M_\odot~\rm pc^{-2}$ and $> 100~M_\odot~\rm pc^{-2}$ are 0.55 and 0.62, respectively.

Here, we compare our results with previous $R_{21}$ measurements toward GMCs in other galaxies. The $R_{21}$ in NGC~1300 is similar to that in the Milky Way \citep{sakamoto_out--plane_1997}. In the local interarm region, where star formation activity is low, the mean $R_{21}$ of the clouds is 0.48, which is comparable to that in {\it bar}. On the other hand, the mean value is 0.68 in the local arm region (including Orion A and B), which is comparable to that in {\it bar-end} and slightly higher than that in {\it arm}. $R_{21}$ of the GMCs in nearby grand-design spiral galaxy NGC~628 is $\sim0.4-0.7$ \citep[mean value of 0.54;][]{herrera_headlight_2020}, which is also comparable to that in NGC~1300. There appears to be no correlation between GMC mass and $R_{21}$ in NGC~628, which is consistent with Figure~\ref{fig:R21 vs. GMC prop}(a). 
Based on the similarity, typical physical conditions of GMCs in NGC~1300 may be similar to those in the Milky Way and NGC~628. However, higher $R_{21}$ ($\geq 1.0$) of GMCs is seen in the central region of the Milky Way \citep[$\sim 1.1$;][]{oka_co_1996}, LMC \citep[$\sim 1.0$;][]{johansson_results_1998,sorai_co_2001}, while such a GMC is not found in NGC~1300 (maximum $R_{21}$ of GMC is 0.80). Such regions with high $R_{21}$ would have significantly different environments from NGC~1300. For example, \citet{sawada_tokyo-onsala-eso-calan_2001} suggest that the high ratio in the center of the Milky Way is due to the lower optical depth and higher excitation of the CO lines ($J \geq 3$) under the higher pressure and temperature.

\section{Discussion}

\subsection{Environmental variations in $R_{21}$ on 100~pc scale}

\begin{figure*}[htbp]
\begin{center}
\includegraphics[width=\hsize]{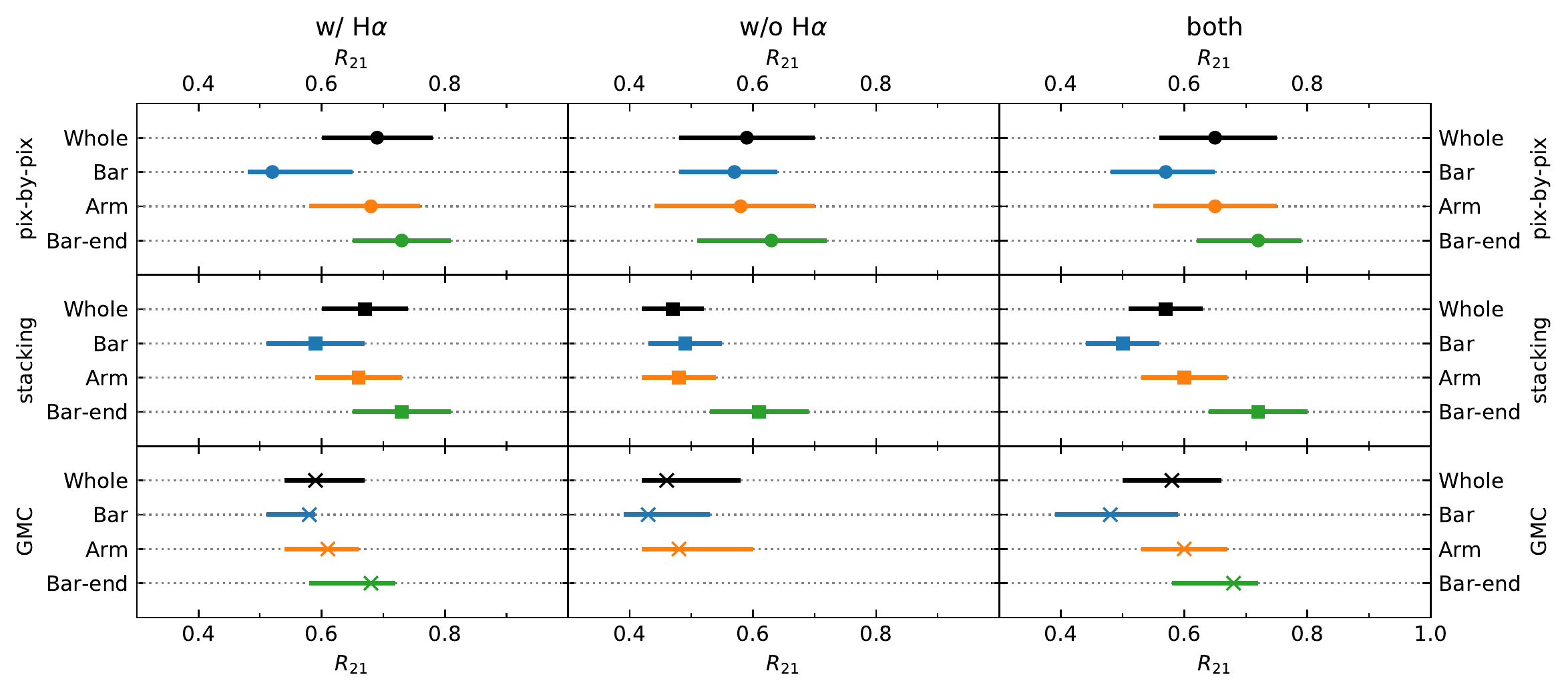}
\caption{Summary of $R_{21}$ on 100~pc scale in NGC~1300. Top, middle, and bottom panels show the results derived by pixel-by-pixel method, stacking analysis, and GMC identification method, respectively. Left and middle columns show $R_{21}$ of pixels (GMCs) with and without H$\alpha$, respectively.
Right column show $R_{21}$ of all pixels (GMCs).
Bars show the IQR in the top and middle panels, while those show the measurement uncertainties for the bottom panel.}
\label{fig:R21 summary}
\end{center}
\end{figure*}

Here, we summarize the results based on the three analyses conducted in this study. Figure~\ref{fig:R21 summary} compares the $R_{21}$ on 100~pc scale in NGC~1300 obtained by each analysis. Bars show the IQR in the panels of the  pixel-by-pixel and GMC identification methods, while those show the measurement uncertainties for the stacking analysis panel. Regardless of the method, the $R_{21}$ tends to be the highest in {\it bar-end}, followed by {\it arm} and {\it bar}.
As mentioned in Section~\ref{sec: Stacking analysis}, the pixel-by-pixel method is likely to overestimate the $R_{21}$, especially where H$\alpha$ is not detected, because the measurement pixels are biased toward the CO(2--1) bright regions. Based on the stacking analysis and GMC identification method, the typical $R_{21}$ in {\it bar-end}, {\it arm}, and {\it bar} are $0.72$, $0.60$, and $0.50$, respectively. 

Our study shows that the $R_{21}$ depends on the presence or absence of the H$\alpha$ (i.e., massive star formation). The pixels (GMCs) with H$\alpha$ have systematically higher $R_{21}$ than those without H$\alpha$. It is reasonable that $R_{21}$ is high in the pixels (GMCs) with H$\alpha$ because the gas density is so high that star formation occurs and the formed OB stars heat the gas there. In {\it bar}, since massive star formation is suppressed \citep{maeda_properties_2020}, H$\alpha$ emission is not associated with most GMCs, resulting in the lowest typical $R_{21}$. The cause for the low star formation activity will be discussed in the next section. The $R_{21}$ of the pixels with and without H$\alpha$ in {\it bar-end}, are higher than those in {\it arm}, respectively, resulting in the highest typical $R_{21}$ in {\it bar-end}. Perhaps the gas in {\it bar-end} is systematically denser than {\it arm} (see next section).

\subsection{Evolution in physical conditions of the molecular gas} \label{sec: evolution in physical conditions of the molecular gas}

In spiral arm structures, changes in $R_{21}$ from low in upstream interarm regions to high in the downstream side of the spiral arms have been reported in some disk galaxies; the Milky Way \citep{sakamoto_out--plane_1997}, M51 \citep{koda_physical_2012}, and M83 \citep{koda_systematic_2020}. These results show the evolution in physical conditions of the gas as it passes through the spiral arms.
In this section, we will examine if such a trend is seen in NGC~1300 using stacking analysis.

\begin{figure}[htbp]
\begin{center}
\includegraphics[width=\hsize]{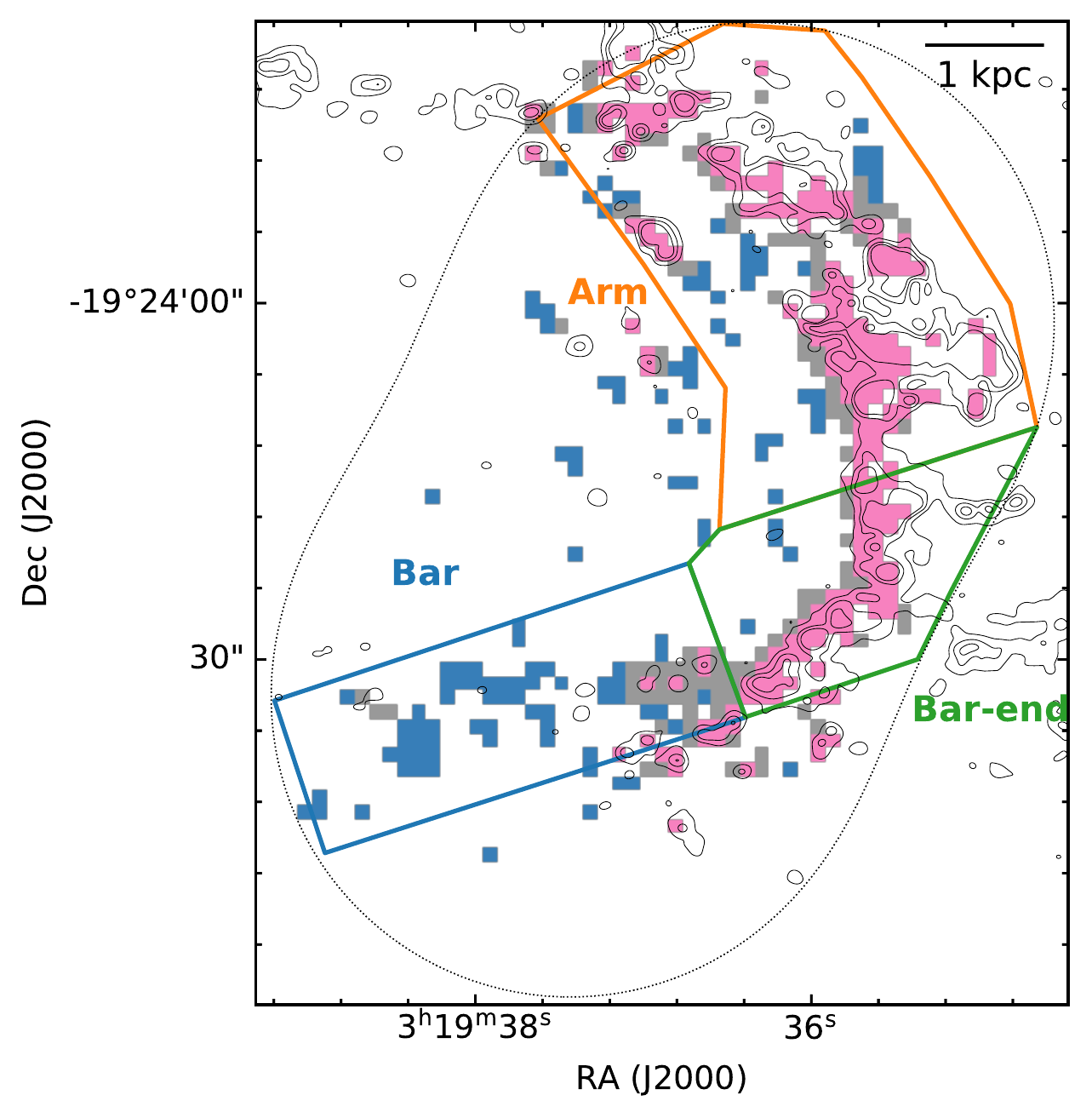}
\caption{Classification of the pixels in which CO(1--0) was detected. Pixel size is 120~pc. Blue pixels show pixels in which no H$\alpha$ was detected and no H$\alpha$ is detected in its surrounding eight pixels (denoted as type A in the text). Gray pixels show 
pixels in which no H$\alpha$ was detected but H$\alpha$ is detected in at least one pixel of its surrounding eight pixels (type B). Pink pixels show pixels in which H$\alpha$ was detected (type C). Black contours show H$\alpha$ emission with 3$\sigma$, 10$\sigma$, and 20$\sigma$, respectively. }
\label{fig: classification}
\end{center}
\end{figure}

First, the pixels with a size of 120~pc in which CO(1--0) was detected are classified into the following three types: (A) pixels in which no H$\alpha$ was detected and no H$\alpha$ is detected in its surrounding eight pixels, (B) pixels in which no H$\alpha$ was detected but H$\alpha$ is detected in at least one pixel of its surrounding eight pixels, and (C) pixels in which H$\alpha$ was detected (same as noted as "w/ H$\alpha$" in Table~\ref{tab:stacking R21 pix-by-pix}). Figure~\ref{fig: classification} shows the distribution of the three types of pixels. 
In {\it arm} and {\it bar-end}, type A pixels (blue pixels) are distributed in the region close to the upstream interarm region.
Type B (gray) and C (pink) pixels are distributed in the spiral arm structure. While Type B pixels are mostly distributed on the upstream edge of the arm, Type C pixels are mostly located downstream from Type B pixels.

We investigate $R_{21}$ for each type by using the same stacking analysis as described in Section~\ref{sec: Stacking analysis}. In {\it arm}, we can see the same changes in $R_{21}$ as in the spiral arm of Milky Way, M51, and M83. The stacked $R_{21}$ of type A, B, and C pixels are $0.38\pm0.05$, $0.57\pm0.07$, and $0.66\pm0.07$, respectively. The $R_{21}$ increases from upstream interarm regions to the downstream side of the spiral arm. Compared to the difference between type A and B (0.19), the difference between type B and C is small (0.09). Based on the previous non-LTE analysis \citep[e.g.,][]{koda_physical_2012}, the molecular gas in type A pixels, where massive star formation is not seen and $R_{21}$ is low, implies that the density and temperature are lower than that in the star-forming regions in the arm structure. The molecular gas of type B pixels is thought to be at the entry to the arm structure. A plausible explanation of higher $R_{21}$ in type B pixels than type A would be that gas density is raised by compression. The temperature may also increase. Newly formed massive stars, which are obscured in their parental GMCs, may exist in the upstream edge of the H$\alpha$ bright region. Surrounding massive star formation may heat the gas in type B pixels. It is reported that $R_{21}$ is high in interfaces of H\textsc{ii} regions in the Milky Way \citep[e.g.,][]{nishimura_revealing_2015}. Although  H\textsc{ii} regions are generally much smaller than  
the pixel size of 120~pc, 
such a situation might be included in type B pixels.
In type C pixels, which overlap with H$\alpha$, young stellar heating would be strong, resulting in higher $R_{21}$. CO(1--0) was not detected in the downstream side of the H$\alpha$ bright region. This would be because molecular gases are dispersed by feedback caused by massive star formation such as photoionization and stellar winds, which are considered as the main mechanisms for dispersing the GMCs \citep[e.g.,][]{kruijssen_fast_2019,chevance_lifecycle_2020}.

In {\it bar-end}, the same trend as in {\it arm} is seen. The stacked $R_{21}$ of type A, B, and C pixels are $0.25\pm0.08$, $0.65\pm0.08$, and $0.73\pm0.08$, respectively. $R_{21}$ of type B and C in {\it bar-end} are higher than those in {\it arm}. Since the $L_{\rm H\alpha}$ is comparable and the surface density of star formation rate derived from FUV and IR is also comparable \citep{maeda_large_2020} in both regions, there may be no difference in the heating effect. Therefore, the gas density in {\it bar-end} may be higher than that in {\it arm}. This is consistent with recent studies that show the presence of the dense gas or high gas density in bar-end regions of the Milky Way and nearby galaxies \citep[e.g.,][]{gallagher_dense_2018,yajima_co_2019,kohno_forest_2020,2021MNRAS.506..963B}. For example, in NGC~3627, \citet{watanabe_3_2019} showed that there is denser molecular gas in bar-end region than in the spiral arm by comparing the chemical composition. \citet{kohno_forest_2020} proposed the converging gas flow from the arm and bar causes the highly turbulent condition, which makes frequent cloud-cloud collisions (CCCs). This scenario can explain the origin of the dense gas formation, which is consistent with some simulations \citep[e.g.,][]{renaud_environmental_2015, takahira_cloud-cloud_2014, takahira_formation_2018}. In fact, \citet{maeda_connection_2021} suggest that high-speed CCCs between massive GMCs occur in {\it bar-end} of NGC~1300, which may form dense molecular gas.

The evolution of molecular gas in {\it bar} is clearly different from those in {\it arm} and {\it bar-end}. Unlike those regions, most of the pixels in {\it bar} are type A, which is located on the dust lane. The stacked $R_{21}$ of type A, B, and C pixels are $0.44\pm0.06$, $0.54\pm0.07$, and $0.59\pm0.08$, respectively. The $R_{21}$ of type A in {\it bar} is higher than those in {\it arm} and {\it bar-end}. This result implies that the molecular gas in the dust lane of the bar region is denser than that in the upstream interarm regions. Nevertheless, star formation activity in {\it bar} of NGC~1300 is clearly suppressed, as evidenced by the small number of type C pixels. Such suppression of massive star formation is seen not only in NGC 1300, but also in  other barred galaxies \citep[e.g.,][]{downes_co_1996,reynaud_kinematics_1998,james_h_2009,maeda_large_2018,maeda_properties_2020,maeda_large_2020}. Some explanations have been proposed as the cause for the star formation suppression of GMCs in the bar regions; the strong shock or/and shear, which would make GMCs stripped, shredded, and finally destroyed  \citep[e.g.,][]{tubbs_inhibition_1982,athanassoula_existence_1992,reynaud_kinematics_1998,emsellem_interplay_2015}, the presence of gravitationally unbound GMCs \citep{sorai_properties_2012, nimori_dense_2013}, and high-speed CCC, which causes too short duration for the cores to grow large enough for massive star formation via gas accretion \citep{fujimoto_environmental_2014,fujimoto_fast_2020}. A recent GMC study of NGC~1300 suggests that high-speed CCC between low-mass GMCs is  the leading candidate of the cause for the suppression \citep{maeda_connection_2021}.
Note that the lower molecular gas mass in {\it bar} would be the cause for the different star formation activity between {\it bar} and {\it bar-end} with the same high CCC speed \citep[see also Sections 5.2 and 5.3;][]{maeda_connection_2021}.

\begin{figure*}[htbp]
\begin{center}
\includegraphics[width=\hsize]{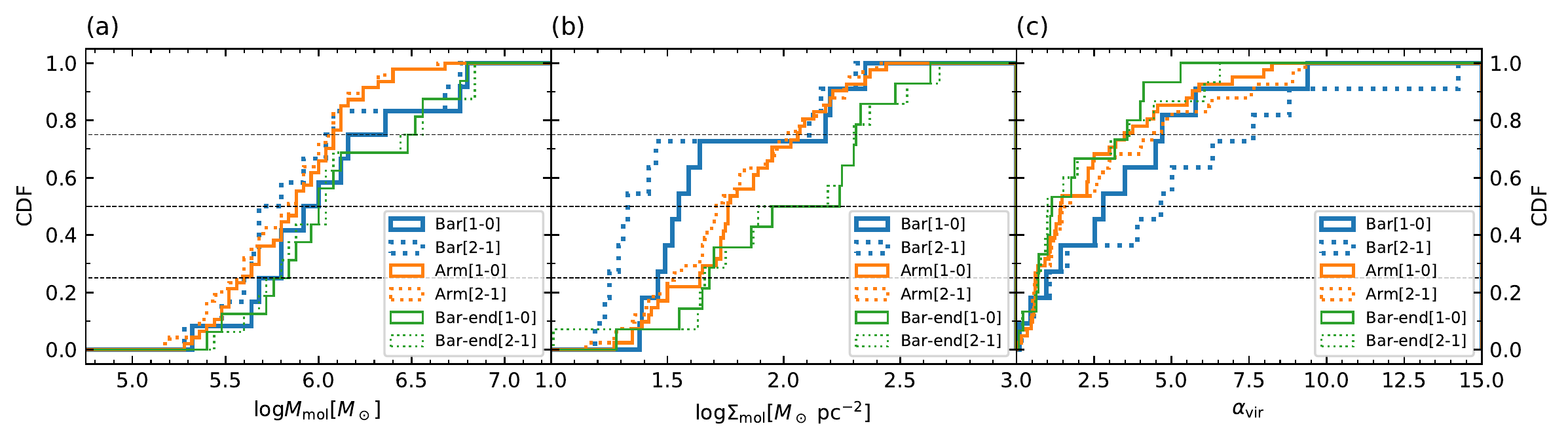}
\caption{ (a) Comparison of the molecular gas mass of the GMCs obtained from CO(1--0) (solid lines) with those obtained from CO(2--1) by assuming the constant $R_{21}$ of 0.65 (dotted lines). (b) Same as panel (a), but for the molecular gas surface density. (c) Same as panel (a), but for the virial parameter.
}
\label{fig:R21 GMC CO10 vs CO21}
\end{center}
\end{figure*}

\subsection{Dependence on spatial scale} \label{sec: Dependence on spatial scale}

\citet{maeda_large_2020} measured the $R_{21}$ in NGC~1300 on an angular resolution of $16.7^{\prime\prime}$, corresponding to 1.67~kpc. The $R_{21}$ in {\it bar}, {\it arm}, and {\it bar-end} are $0.26 \pm 0.02$, $0.52 \pm 0.03$, and $0.66 \pm 0.03$.  These values are lower than those on $\sim 100$ pc scale and the difference is largest in {\it bar}; the difference from  the stacking analysis (GMC identification method) is 0.24 (0.22), 0.08 (0.08), and 0.06 (0.02) in {\it bar}, followed by {\it arm} and {\it bar-end}, respectively.

The cause for the low $R_{21}$ on kpc scale would be the presence of an extended diffuse molecular gas, which is distributed on scales larger than sub-kpc and would not contribute to the star formation activity. 
\citet{maeda_large_2020} measured diffuse molecular gas fraction,  which is derived using the CO(1--0) flux obtained from ALMA 12-m array, which has no sensitivity on diffuse (extended; full width at half-maximum $\geq 700$~pc) molecular gases due to the lack of ACA, and the total CO(1--0) flux obtained from NRO 45-m telescope.
They showed that the diffuse molecular gas fraction is $0.74-0.91$ in {\it bar} and $0.28-0.65$ in {\it arm} and {\it bar-end}. This indicates most of the molecular gas in {\it bar} exists as  diffuse gas. Furthermore, they find a tight anti-correlation of $R_{21}$ and the ratio of the diffuse gas, suggesting the $R_{21}$ of the diffuse gas is low. In our data cube with a high spatial resolution of $\sim100$~pc, it is thought that such diffuse component is too faint to be detected (even if it is detected, the GMC identification method would exclude the diffuse gas). Therefore, the pixels we used to measure $R_{21}$ in this study would not contain the majority of the diffuse gas, resulting in a higher $R_{21}$ than that measured on kpc scale. In fact, the stacked $R_{21}$ by using all pixels in {\it bar} is roughly comparable to that on kpc scale; By predicting the $\bar{v}_{\rm CO(1-0)}$ in pixels where CO was not detected based on a velocity-field model \citep[][see Figure~2(b)]{maeda_connection_2021}, stacked $R_{21}$ in {\it bar} is calculated to be $0.33 \pm 0.04$. This value is lower than that derived from the stacking analysis by using only CO(1--0) detected pixels (Section~\ref{sec: Stacking analysis}), and is roughly comparable to that measured by \citet{maeda_large_2020},
which supports the idea that the majority of diffuse CO component was not detected in our data cube with $\sim 100$~pc resolution.

Such difference between $R_{21}$ of the GMCs and that measured on kpc scale may commonly be seen in disk galaxies because a certain amount of the diffuse molecular gas is expected to be universally exist in the disk galaxies as reported by some studies \citep[e.g.,][]{pety_plateau_2013,caldu-primo_spatially_2015}. We emphasize that the difference would be large in regions with a large amount of diffuse molecular gas like {\it bar} in NGC~1300.

\subsection{Impact of  using a constant $R_{21}$} \label{sec: Impact of  using a constant R21}

Molecular gas mass of a GMC is usually converted from the CO(1--0) luminosity, $L_{\rm CO(1-0)}$ by adopting a $\alpha_{\rm CO}$:
\begin{equation}
    \left[ \frac{M_{\rm mol}^{10}}{M_\odot} \right] = \alpha_{\rm CO} \left[ \frac{L_{\rm CO(1-0)}}{\rm K~km~s^{-1}~pc^2} \right].
\end{equation}
In this study, we measured $M_{\rm mol}^{10}$ by assuming $\alpha_{\rm CO} = 4.4~M_\odot~\rm (K~km~s^{-1}~pc^2)^{-1}$ \citep{bolatto_co--h2_2013}.
On the other hand, when converting the mass from CO(2--1) luminosity, the line ratio is often assumed to be constant, $R_{21}^{\rm const}$:
\begin{equation}
    \left[ \frac{M_{\rm mol}^{\rm 21}}{M_\odot} \right] = \alpha_{\rm CO} \frac{1}{R_{21}^{\rm const}}\left[ \frac{L_{\rm CO(2-1)}}{\rm K~km~s^{-1}~pc^2} \right].
\end{equation}
The $R_{21}^{\rm const}$ of 0.65--0.7 is usually assumed in recent GMC studies \citep[e.g.,][]{wu_submillimeter_2017, sun_cloud-scale_2018, rosolowsky_giant_2021}. However, $R_{21}$ of the GMCs is different among environments and has a certain amount of scatter as described in Section~\ref{sec: R21 of GMCs}. Therefore, the assumption of the constant $R_{21}$ would affect the discussion of the differences in GMC properties among environments. Here, we see the impact.
To derive $M_{\rm mol}^{\rm 21}$, we convert from $L_{\rm CO(1-0)}$ to $L_{\rm CO(2-1)}$ using the $R_{21}$ derived in Section~\ref{sec: R21 of GMCs}. Then, we derived $M_{\rm mol}^{\rm 21}$ using $R_{21}^{\rm const}$ of 0.65.
In the same way, we derived molecular gas surface density, $\Sigma_{\rm mol}^{\rm 21}$, and virial parameter, $\alpha_{\rm vir}^{\rm 21}$. The virial parameter is a useful measure of the gravitational binding and is defined as $5\sigma_v^2R/G M_{\rm mol}$ by \citet{bertoldi_pressure-confined_1992}. 

Figure~\ref{fig:R21 GMC CO10 vs CO21} compares the  physical properties obtained from CO(1--0) ($M_{\rm mol}^{\rm 10}$, $\Sigma_{\rm mol}^{\rm 10}$, and $\alpha_{\rm vir}^{\rm 10}$) with those obtained from CO(2--1) by assuming a constant $R_{21}^{\rm const}$ of 0.65. Because the $R_{21}$ in {\it arm} and {\it bar-end} is close to 0.65, there is little difference in the cumulative distributions. On the other hand, because the $R_{21}$ in {\it bar} is systematically lower than 0.65, physical properties obtained from CO(2--1) are systematically different by $40 \sim 70~\%$ compared to those from CO(1--0): the median values are $(M_{\rm mol}^{\rm 10}, M_{\rm mol}^{\rm 21}) = (1.0 \times 10^6~M_\odot, 5.9 \times 10^5~M_\odot)$, $(\Sigma_{\rm mol}^{\rm 10}, \Sigma_{\rm mol}^{\rm 21}) = (35.5~M_\odot~{\rm pc}^{-2}, 21.7~M_\odot~{\rm pc}^{-2})$, and $(\alpha_{\rm vir}^{\rm 10}, \alpha_{\rm vir}^{\rm 21}) = (2.8, 4.7)$, respectively.  Such under- or overestimations would lead to wrong discussion on difference among environments:
For example, although the environmental difference in $\alpha_{\rm vir}^{\rm 10}$ is not seen, the $\alpha_{\rm vir}^{\rm 21}$ in {\it bar} seems to be higher than those in {\it arm} and {\it bar-end}. In fact, the $p_{\rm KS}$ for {\it bar} versus {\it arm} when using $\alpha_{\rm vir}^{\rm 21}$ (0.13) is smaller than when using $\alpha_{\rm vir}^{\rm 10}$ (0.44).
Therefore, when comparing the properties of GMCs between environments with low star formation activity and low $R_{21}$ (e.g., bar, interarm) and opposite environments (e.g., spiral arm, bar-end), it is not desirable to assume a constant $R_{21}$.

It is important to note that there are other factors that can cause systematic biases in the measurement of GMC properties.
There is a possibility that the $\alpha_{\rm CO}$ changes among environments. \citet{sorai_properties_2012} suggest that $\alpha_{\rm CO}$ in the bar regions may be $0.5-0.8$ times smaller than that in the arm regions in Maffei 2 using large velocity gradient analysis. Similar possibilities are pointed out \citep[][]{morokuma-matsui_stacking_2015, watanabe_refined_2011}. Thus, molecular gas mass in {\it bar} may be overestimated. Further, measured GMC properties depend on spatial resolution. Cloud identification algorithm tends to divide CO emission into structures with a rather uniform size scale, comparable to the spatial resolution \citep{hughes_comparative_2013}. In fact, the median radius of the GMCs in NGC~1300 is 48~pc using the CO(1--0) cube with 40~pc resolution \citep{maeda_properties_2020}, while the median is 105 pc in this study using $\sim$100~pc resolution data cube. This result suggests that the GMCs we identified are likely to be blends of smaller GMCs (in fact, the number of GMCs is approximately halved).  Thus, Figure~\ref{fig:R21 GMC CO10 vs CO21} would show the properties of the aggregation of multiple GMCs. To measure $R_{21}$ of GMCs more accurately, it is desirable to observe CO(2--1) with a higher spatial resolution ($\sim 40$~pc), which is a future task.

\section{Summary}

We measured the brightness temperature ratio of $R_{21} = ^{12}{\rm CO}(2-1)/^{12}{\rm CO}(1-0)$ in the strongly barred galaxy NGC~1300 on a spatial scale of $\sim$100~pc. We observed the CO(1--0) emission using ALMA and NRO 45-m telescope, and CO(2--1) emission was obtained from ALMA archival data. We measured the ratio not only on a pixel-by-pixel basis but also by identifying the GMCs using 3D clumps finding algorithm. In pixel-by-pixel measurements, we performed a stacking analysis.
We investigated the environmental dependence of the $R_{21}$ and the relationship between the ratio and star formation activity using a continuum-subtracted H$\alpha$ image. The main results are as follows:

\begin{enumerate}
    \item Regardless of the method, the $R_{21}$ tends to be the highest in {\it bar-end}, followed by {\it arm} and {\it bar}. The pixels (GMCs) with H$\alpha$ have a systematically higher ratio than those without H$\alpha$ (Figure~\ref{fig:R21 summary}).
    \begin{itemize}
        \item Based on the pixel-by-pixel analysis, which used the pixels where both CO(1--0) and CO(2--1) were detected, the resultant median $R_{21}$ is 0.65 with a scatter (i.e. IQR) of 0.19 in the whole region (Figure~\ref{fig:R21 pix-by-pix hist} and Table~\ref{tab:R21 pix-by-pix}). We find a moderate correlation between the $R_{21}$ and the H$\alpha$ luminosity: $R_{21} \propto L_{\rm H\alpha}^{0.13 \pm 0.20}$ (Figure~\ref{fig:R21_vs_Ha}).

        \item Stacking with all pixels where CO(1--0) was detected, we find that the pixel-by-pixel method is likely to overestimate the median $R_{21}$ of the pixels without H$\alpha$. The stacked $R_{21}$ in the whole region, {\it bar-end}, {\it arm}, and {\it bar} are $0.57 \pm 0.06$, $0.72 \pm 0.08$, $0.60 \pm 0.07$, and $0.50 \pm 0.06$, respectively.  In the whole region, the stacked $R_{21}$ of the pixels with H$\alpha$ of $0.67 \pm 0.07$ is higher than those without H$\alpha$ of $0.47 \pm 0.05$ (Table~\ref{tab:stacking R21 pix-by-pix}). 
        
        \item The $R_{21}$ of the GMCs, which were identified by \textsc{cprops}, is consistent with the ratio by the stacking analysis (Figure~\ref{fig:R21 GMC hist} and Table~\ref{tab:R21 GMC}).
        The $R_{21}$ tends to slightly decrease with increasing GMC radius, and there is a moderate correlation for $R_{21}$ vs. molecular gas surface density (Figure~\ref{fig:R21 vs. GMC prop}).

        \item In {\it bar}, since massive star formation is suppressed, H$\alpha$ emission is not associated with most of the pixels (GMCs), resulting in the lowest $R_{21}$.
        
    \end{itemize}

    \item The $R_{21}$ in NGC~1300 changes from low in the upstream interarm regions to high in downstream side of the spiral arm, which is evidence of the evolution in physical conditions of the gas as it passes through the spiral arms (Section~\ref{sec: evolution in physical conditions of the molecular gas}).

    \item The $R_{21}$ measured on 1.67~kpc scale tends to be lower than that on $\sim 100$ pc scale. The cause of this difference is the presence of an extended diffuse molecular gas, which is distributed on scales larger than sub-kpc and would not be detected in the image with $\sim 100$ pc resolution due to the low sensitivity. The difference between the two ratios is the largest in {\it bar} where the diffuse molecular gas fraction is the highest in NGC~1300 (Section~\ref{sec: Dependence on spatial scale}).

    \item The assumption of constant $R_{21} = 0.65$, which is usually used in recent GMC studies by CO(2--1) observations, tends to systematically over- or underestimate the properties of GMCs (i.e., molecular gas mass, surface density, and virial parameter)  in regions where star formation activity is low (i.e., {\it bar}). This assumption would
    lead to different conclusions on environmental variations in the properties of the GMCs than CO(1--0) observations.
    
\end{enumerate}

\begin{acknowledgments}
FM is supported by Research Fellowship for Young Scientists from the Japan Society of the Promotion of Science (JSPS). FE is supported by JSPS KAKENHI Grant Number JP17K14259. KO is supported by JSPS KAKENHI Grant Number JP19K03928. AH is supported by the JSPS KAKENHI Grant Number JP19K03923. The Nobeyama 45-m radio telescope is operated by NRO, a branch of National Astronomical Observatory of Japan (NAOJ). This paper makes use of the following ALMA data: ADS/JAO.ALMA \#2015.1.00925.S. \#2017.1.00248.S. \#2018.1.01651.S \#2019.2.00139.S. ALMA is a partnership of ESO (representing its member states), NSF (USA), and NINS (Japan), together with NRC (Canada), MOST and ASIAA (Taiwan), and KASI (Republic of Korea), in cooperation with the Republic of Chile. The Joint ALMA Observatory is operated by ESO, AUI/NRAO, and NAOJ. Data analysis was in part carried out on the Multi-wavelength Data Analysis System operated by the Astronomy Data Center (ADC), NAOJ.
\end{acknowledgments}


\vspace{5mm}
\facilities{ALMA, HST, NRO:45m}


\software{CASA \citep[ver. 5.1.1., 5.6.1., and 5.7.2.][]{mcmullin_casa_ASPCS}, NOSTAR \citep{sawada_onthefly_2008}, Astropy \citep{astropy_2018}, APLpy \citep{Robitaille_2012}}



%\appendix


\bibliographystyle{aasjournal}
\bibliography{Reference_R21}



\end{document}